\documentclass[12pt]{article}
\usepackage{epsfig}
\begin{document}
\title{Probing the structure of $X(3872)$ in photoproduction}
\author{E. Ya. Paryev\\
{\it Institute for Nuclear Research of the Russian Academy of Sciences}\\
{\it Moscow, Russia}}

\renewcommand{\today}{}
\maketitle

\begin{abstract}
We study the production of $X(3872)$ mesons in photon-induced nuclear
reactions near the threshold within the collision model based on the nuclear spectral function. The model
accounts for direct photon-nucleon $X(3872)$ production processes as well as five different scenarios for their internal structure. We calculate the absolute and relative excitation functions for $X(3872)$ production off $^{12}$C and $^{184}$W target nuclei at near-threshold incident photon energies of 8--16 GeV, the absolute differential cross sections for their production off these target nuclei at laboratory angles of 0$^{\circ}$--10$^{\circ}$ and for incident photon energy of 13 GeV as well as the A dependences of the relative (transparency ratios) cross sections for $X(3872)$ production from ${\gamma}A$ collisions at photon energies around 13 GeV within the adopted scenarios for the $X(3872)$ meson internal structure. We show that the absolute and relative observables considered reveal distinct sensitivity to these scenarios. Therefore, the measurement of such observables in a dedicated experiment at the CEBAF facility
in the near-threshold energy range will allow us to get valuable information on the $X(3872)$ inner structure.
\end{abstract}

\newpage

\section*{1. Introduction}

\hspace{1.5cm} The discovery of $X(3872)$ resonance (also known as $\chi_{c1}(3872)$) by the Belle Collaboration in
2003 [1] as a narrow peak in the vicinity of the $D^0{\bar D}^{*0}$ (${\bar D}^0{D}^{*0}$) mass threshold in the ${J/\psi}{\pi^+}\pi^-$ invariant mass distribution in exclusive $B^{\pm} \to K^{\pm}X(3872) \to K^{\pm}({J/\psi}{\pi^+}\pi^-)$ decays
\footnote{$^)$It was also confirmed in many other high-energy experiments. In particular, the experimental results of studying the $X(3872)$ lineshape using the data samples collected with the Belle and BESIII detectors have been reported in very recent publications [2] and [3], respectively.}$^)$
has opened a new era for the study of exotic heavy hadrons, which exhibit properties incompatible with the
predictions of the traditional quark model for quark-antiquark mesons and three-quark baryons.
They are composed of four or five quarks (and antiquarks) and usually are named as tetraquark and pentaquark
states, respectively. Since then, many unconventional charmonium- and bottomonium-like states, the so-called
$X, Y, Z$ mesons as well as the hidden-charm non-strange $P_c$ and strange $P_{cs}$ pentaquark states,
the doubly-charmed tetraquark $T^+_{cc}(3875)$ state, the fully charmed tetraquark $X(6900)$ state
and so on have been observed in various experiments, as summarized in reviews [4--19]. The observation of such
exotic hadronic states and studying their properties are very important for uderstanding the nonperturbative aspects of QCD and for extending our knowledge about the world of "elementary" particles. Among these exotic hadronic states, the $X(3872)$ attracted lots of attention because of its rather enigmatic intrinsic structure.
Since the experimental discovery of the $X(3872)$ meson, this structure has been intensely debated.
However, a compelling understanding of the nature of this resonance is still lacking and there is no consensus
about its structure. Thus, in a molecular scenario, due to the closeness of the observed
$X(3872)$ mass ($m_{X(3872)}=3871.65\pm0.06$ MeV [20]) to the  $D^0{\bar D}^{*0}$ (${\bar D} ^0{D}^{*0}$) mass
threshold ($m_{D^0}+m_{{\bar D}^{*0}}=3871.69\pm$0.07 MeV), the $X(3872)$ resonance can be naturally interpreted as an extremely narrow $D^0{\bar D}^{*0}$+${\bar D} ^0{D}^{*0}$ molecular state in a relative S-wave with a width less than 1.2 MeV [20] and with a very small binding energy (see, for instance, Refs. [21, 22] and those given below).
However, there exists another alternative explanations of the $X(3872)$. For example, it is interpreted as a conventional $c{\bar c}$ charmonium $\chi_{c1}(2P)$ state [23], a diquark-antidiquark compact tetraquark state [24], a mixture of a molecule and an excited charmonium state [25]. Such also plausible interpretations of the $X(3872)$ exotic state were supported by many current theoretical studies (see references herein below).

Most of the above interpretations are based on the comparison of the calculated and experimental branching fractions for
various two- and three-body $X(3872)$ decays. However, the internal structure of $X(3872)$ could also be investigated in high-energy proton--proton and heavy--ion collisions at hadron colliders, since its production yield in these collisions could reflect its structure. Thus, the rather large prompt ({\it i.e.} not from $B$ decays) production rate of $X(3872)$ at hadron colliders is considered as an argument against its interpretation as a weakly bound charm-meson
molecule (cf. [26]). There are many debates in the literature [26--32] on whether or not it is compatible with such interpretation, leading to mutually conflicting conclusions. Recently, the inclusive high transverse momentum distributions of $X(3872)$, promptly produced in high-energy $pp$ [33, 34, 35] and PbPb [36] collisions
at rates commensurate with those of the $\psi(2S)$ charmonium, were measured at the LHC, respectively,
by the CMS, ATLAS, LHCb and CMS Collaborations. A similarity of the $X(3872)$ production rate with the $\psi(2S)$
production rate suggests  the importance  of a compact component in the X(3872) wave function. Indeed, the measured
in $pp$ collisions $X(3872)$ transverse momentum distributions are well described in Refs. [37, 38] assuming a large
charmonium $c{\bar c}$ component in this function. Nevertheless, in spite of large efforts the full understanding the production of $X(3872)$ and its structure in hadronic collisions remains a challenging open problem [13, 16, 18, 19, 39]. Alternatively, the structure of $X(3872)$ mesons can also be studied in their photoproduction off nuclei at energies
close to the threshold for their production off a free nucleon. This has the advantage compared to the hadronic collisions that the interpretation of data from such experiments is less ambiguous owing to a negligible strength of
initial-state photon interaction and since the $X(3872)$ meson production occurs through a few channels in a cleaner environment - in static cold nuclear matter whose density is well known.

In this paper we present the detailed results for the absolute and relative excitation functions for $X(3872)$ production off $^{12}$C and $^{184}$W target nuclei as well as for the A dependences of the relative (transparency ratios) cross sections for $X(3872)$ production from ${\gamma}A$ collisions in the threshold energy region obtained in the framework of the first collision model, based on the nuclear spectral function, within the different scenarios for the $X(3872)$ meson internal structure. A comparison of these results with data, which could be taken in the future JLab experiments at the CEBAF facility, can provide deeper insights into the inner structure and properties of the puzzling $X(3872)$ resonance.

\section*{2. The model: direct $X(3872)$ photoproduction mechanism}

\hspace{1.5cm} Direct $X(3872)$ photoproduction on nuclear targets in the near-threshold incident laboratory photon energy region $E_{\gamma} \le 16$ GeV
\footnote{$^)$Which corresponds to the center-of-mass energies W of the photon-proton system $W \le 5.56$ GeV,
or to the relatively "low" excess energies $\epsilon_{X(3872)p}$ above the $X(3872)p$ production threshold
$W_{\rm th}=m_{X(3872)}+m_p=$4.81 GeV ($m_{X(3872)}$ and $m_p$ are the $X(3872)$ meson and proton bare masses,
respectively ) $0 \le \epsilon_{X(3872)p} \le 0.75$ GeV and in which the exotic hadron $X(3872)$ can be observed
in ${\gamma}p$ and ${\gamma}A$ reactions at the 12 GeV CEBAF facility and at the proposed 24 GeV apgrade of this
facility with novel magnet designs in the existing recirculation arcs [40--44].}$^)$
may proceed via the following elementary processes with the lowest free production threshold ($\approx$ 11.86 GeV) [45--47]:
\begin{equation}
{\gamma}+p \to X(3872)+p,
\end{equation}
\begin{equation}
{\gamma}+n \to X(3872)+n.
\end{equation}
The $X(3872)$ mesons and nucleons, produced in these processes, are sufficiently energetic.
Thus, for example, the kinematically allowed $X(3872)$ and final proton laboratory momenta in the direct
process (1), taking place on the free target proton at rest, vary within the momentum ranges of
8.481--11.775 GeV/c and 1.225--4.519 GeV/c, respectively, at initial photon beam energy of $E_{\gamma}=13$ GeV.
Since the neutron mass is approximately equal to the proton mass, the kinematical characteristics of
the $X(3872)$ mesons and final neutrons, produced in the reaction (2), are close to those of
final particles ($X(3872)$ mesons and protons) in the process (1).
Evidently, the binding of target nucleons and their Fermi motion will distort the distributions of the
outgoing high-momentum $X(3872)$ mesons and nucleons as well as lead to a wider accessible
momentum intervals compared to those given above.
Since the medium effects are expected to be reduced for high momenta [48--50], we will neglect the modification
of the final high-momentum $X(3872)$ mesons and outgoing nucleons in the nuclear medium in the case when
the reactions (1), (2) proceed on a nucleons embedded in a nuclear target
\footnote{$^)$It should be pointed out that the behavior of $X(3872)$ in a nuclear environment has been studied
in the recent works [51, 52]. The authors of Ref. [51], analyzing the $D{\bar D}^*$ scattering $T$--matrix within
the mixed-molecular scenario of $X(3872)$ have found that the width of the $X(3872)$ peak significantly grows
when the nuclear density is increased, whereas its position moves to higher energies as the molecular component
is lowered. For this component of the order of 60\% for the $X(3872)$, this work predicts at threshold widths
for the resonance of around 30--40 MeV, and more modest repulsive mass-shifts with a maximum of 10 MeV for the nuclear matter saturation density $\rho_0$, which amounts approximately to 0.25\% of its free-space nominal mass.
On the other hand, in Ref. [52] it was obtained using QCD sum rule calculations
based on a diquark-antidiquark picture for the $X(3872)$ resonance that its mass is substantially influenced by the nuclear medium. Indeed, in the approach of Ref. [52] the mass shift of the low-momentum $X(3872)$
due to the nuclear matter is negative and is about 25\% at density $\rho_0$, which is questionable accounting for the
quark content of the $X(3872)$. But in any case in accordance with the above, one may expect that the $X(3872)$-nucleus optical potential should vanish at high $X(3872)$ momenta of interest. The latter is also true for the nucleon--nucleus mean-field potential at high nucleon momenta (cf. Refs. [53, 54]).}$^)$
.

  Then, ignoring the distortion of the incident photon in the energy range of interest
and describing the $X(3872)$ meson final-state absorption by the absorption cross section
$\sigma_{{X(3872)}N}$, we represent the total cross section for the production of $X(3872)$ mesons
on nuclei in the direct photon--induced reaction channels (1), (2) as follows [55, 56]:
\begin{equation}
\sigma_{{\gamma}A\to {X(3872)}X}^{({\rm dir})}(E_{\gamma})=I_{V}[A,\sigma_{{X(3872)}N}]
\left<\sigma_{{\gamma}p \to {X(3872)}p}(E_{\gamma})\right>_A,
\end{equation}
where
\begin{equation}
I_{V}[A,\sigma]=2{\pi}\int\limits_{0}^{R}r_{\bot}dr_{\bot}
\int\limits_{-\sqrt{R^2-r_{\bot}^2}}^{\sqrt{R^2-r_{\bot}^2}}dz
\rho(\sqrt{r_{\bot}^2+z^2})
\exp{\left[-\sigma\int\limits_{z}^{\sqrt{R^2-r_{\bot}^2}}
\rho(\sqrt{r_{\bot}^2+x^2})dx\right]},
\end{equation}
\begin{equation}
\rho(r)=\rho_p(r)+\rho_n(r),\,\,\,r=\sqrt{r_{\bot}^2+z^2}\,\,\, {\rm or}~r=\sqrt{r_{\bot}^2+x^2};
\end{equation}
\begin{equation}
\left<\sigma_{{\gamma}p \to {X(3872)}p}(E_{\gamma})\right>_A=
\int\int
P_A({\bf p}_t,E)d{\bf p}_tdE
\sigma_{{\gamma}p \to {X(3872)}p}(\sqrt{s^*})
\end{equation}
and
\begin{equation}
  s^*=(E_{\gamma}+E_t)^2-({\bf p}_{\gamma}+{\bf p}_t)^2,
\end{equation}
\begin{equation}
   E_t=M_A-\sqrt{(-{\bf p}_t)^2+(M_{A}-m_{p}+E)^{2}}.
\end{equation}
Here, $\sigma_{{\gamma}p\to {X(3872)}p}(\sqrt{s^*})$ is the "in-medium"
total cross section for the production of $X(3872)$ mesons in process (1)
\footnote{$^)$In equation (3) it is assumed that the $X(3872)$ meson production cross sections
in ${\gamma}p$ and ${\gamma}n$ interactions are the same.}$^)$
at the "in-medium" ${\gamma}p$ center-of-mass energy $\sqrt{s^*}$;
$\rho_p(r)$,  $\rho_n(r)$ ($r$ is the distance from the nucleus center)
and $P_A({\bf p}_t,E)$ are normalized to the numbers of protons $Z$, neutrons
$N$ and to unity the local proton, neutron densities and the
spectral function of target nucleus with mass number $A$ ($A=Z+N$), having mass $M_A$ and radius $R$
\footnote{$^)$The specific information about the latter quantity, used in our calculations,
is given in Ref. [57]. The local nucleon densities, employed in the calculations, will be defined below.}$^)$;
${\bf p}_{t}$  and $E$ are the internal momentum and binding energy of the struck target proton
just before the collision; ${\bf p}_{\gamma}$ and $E_{\gamma}$ are the momentum and energy of the incident photon beam.

As in Ref. [57], we suggest that the "in-medium" cross section
$\sigma_{{\gamma}p \to X(3872)p}({\sqrt{s^*}})$ for $X(3872)$ production in process (1)
is equivalent to the vacuum cross section $\sigma_{{\gamma}p \to {X(3872)}p}({\sqrt{s(E_{\gamma})}})$, in which
the vacuum center-of-mass energy squared $s(E_{\gamma})$ for given photon energy $E_{\gamma}$, presented by the formula
\begin{equation}
s(E_{\gamma})=W^2=(E_{\gamma}+m_p)^2-{\bf p}_{\gamma}^2=m_p^2+2m_pE_{\gamma},
\end{equation}
is replaced by the in-medium expression (7).
For the free total cross section $\sigma_{{\gamma}p \to {X(3872)}p}({\sqrt{s(E_{\gamma})}})$
in the considered photon energy range $W \le $ 5.56 GeV we have used the following fit of the results of the calculations of this cross section here within the approach [47], in which the ${\gamma}p \to {X(3872)}p$ reaction is considered to be proceeded by the $t$-channel vector mesons ($\rho$, $\omega$ and $\phi$) exchange:
\begin{equation}
\sigma_{{\gamma}p \to {X(3872)}p}({\sqrt{s(E_{\gamma})}})=125.0\left(1-\frac{s_{\rm th}}{s(E_{\gamma})}\right)^{1.2}~[{\rm nb}],
\end{equation}
where
\begin{equation}
  s_{\rm th}=W_{\rm th}^2=(m_{X(3872)}+m_p)^2.
\end{equation}
In this low-energy regime, the cross sections for the $X(3872)$ production in ${\gamma}p$ collisions are
predicted to be of the order of 1--20 nanobarns [47]. It should be pointed out that in the very
recent work [58] the total cross section of the ${\gamma}p \to X(3872)p$ reaction has been also estimated assuming
the coupled-channel production mechanism through the open-charm meson-baryon intermediate states. For energies
near the threshold, this cross section is predicted to be of the order of tens of nanobarns, which are generally
larger than the model prediction [47] and which together with this prediction are well experimentally measurable
already within the capabilities of the present CEBAF facility (cf. Refs. [59--61]).
At the same time, the authors of this work have underscored that their results provide merely an order-of-magnitude estimate and should not be considered as a precise quantitative prediction since no
data are available so far near the threshold to constrain the model parameters
\footnote{$^)$It should be pointed out that virtual exclusive photoproduction of the $X(3872)$  on a nucleon
target with a muon beam of 160 and 200 GeV/c momentum has been recently explored by the COMPASS Collaboration [62].
A new charmonium-like structure, denoted by the ${\tilde X}(3872)$, has been observed in the
${J/\psi}\pi^+\pi^-$ mass distribution with a 4.1 $\sigma$ statistical significance. Its mass and width are consistent
with those of the $X(3872)$. But the measured $\pi^+\pi^-$ mass distribution from the observed decay into the
${J/\psi}\pi^+\pi^-$ was found to be different from that measured in previous experiments for the $X(3872)$.
This suggests that this structure has the quantum numbers $J^{PC}=1^{+-}$, while the quantum numbers previously
determined for the $X(3872)$ are $1^{++}$. The product of total cross section and branching ratio of the decay
of the observed ${\tilde X}(3872)$ state into the ${J/\psi}\pi^+\pi^-$ was determined to be
71$\pm$28(stat)$\pm$39(syst) pb. Also, the COMPASS Collaboration has measured the upper bound
for the $X(3872)$ photoproduction cross section at substantially higher energy -- at an average ${\gamma}p$
c.m. energy of 13.7 GeV as $\sigma_{{\gamma}N \to X(3872)N'}$$\times$$Br[X(3872) \to {J/\psi}{\pi}\pi] < 2.9$ pb
(CL=90\%).}$^)$
. Therefore, in our subsequent
calculations we will use  the empirical formulas (10), (11) as a guideline for a reasonable estimation of the
$X(3872)$ yield from ${\gamma}A$ interactions.

We focus now on the local nucleon densities, used in our calculations of the $X(3872)$ production on the considered
in the present work target nuclei $^{12}_{6}$C, $^{27}_{13}$Al, $^{40}_{20}$Ca, $^{63}_{29}$Cu, $^{93}_{41}$Nb, $^{112}_{50}$Sn, $^{184}_{74}$W, $^{208}_{82}$Pb and $^{238}_{92}$U. For lightest nucleus $^{12}_{6}$C we use the same proton and neutron density profiles of the harmonic oscillator model [63]. For nuclei $^{27}_{13}$Al, $^{40}_{20}$Ca and $^{63}_{29}$Cu for the proton and neutron densities, $\rho_p(r)$ and $\rho_n(r)$, we have employed in our present calculations the Woods-Saxon distributions with the same radial parameters for protons and neutrons [63], viz.:
\begin{equation}
\rho_p(r)=Z\rho(r),\,\,\,\rho_n(r)=N\rho(r),\,\,\,\rho(r)=\rho_0\left[1+\exp\left(\frac{r-R_{1/2}}{a}\right)\right]^{-1},  \end{equation}
where $\rho(r)$ represents the local nucleon density normalized to unity ($\int\rho(r)d^3r=1$)
and parameters $R_{1/2}$ and $a$, determined from fits to electron-scattering data, are taken from standard compilation [64]. They are: $R_{1/2}=3.07$ fm, $a=0.519$ fm for $^{27}_{13}$Al; $R_{1/2}=3.51$ fm, $a=0.563$ fm for $^{40}_{20}$Ca and $R_{1/2}=4.214$ fm, $a=0.586$ fm for $^{63}_{29}$Cu. For medium-weight $^{93}_{41}$Nb, $^{112}_{50}$Sn and heavy
$^{184}_{74}$W, $^{208}_{82}$Pb, $^{238}_{92}$U target nuclei, we adopted for protons and neutrons the two-parameter Fermi density distributions:
\begin{equation}
\rho_i(r)=\rho_{0i}\left[1+\exp\left(\frac{r-c_i}{z_i}\right)\right]^{-1},\,\,\,i=p,n
\end{equation}
with proton density parameters: $c_{p}=4.87$ fm, $z_p=0.573$ fm for $^{93}_{41}$Nb; $c_{p}=5.375$ fm, $z_p=0.560$ fm
for $^{112}_{50}$Sn and $c_{p}=6.51$ fm, $z_p=0.535$ fm for $^{184}_{74}$W; $c_{p}=6.624$ fm, $z_p=0.549$ fm for $^{208}_{82}$Pb; $c_{p}=6.805$ fm, $z_p=0.605$ fm for $^{238}_{92}$U, also inferred from nuclear charge distributions
for these target nuclei [64].
\begin{figure}[htb]
\begin{center}
\includegraphics[width=15.0cm]{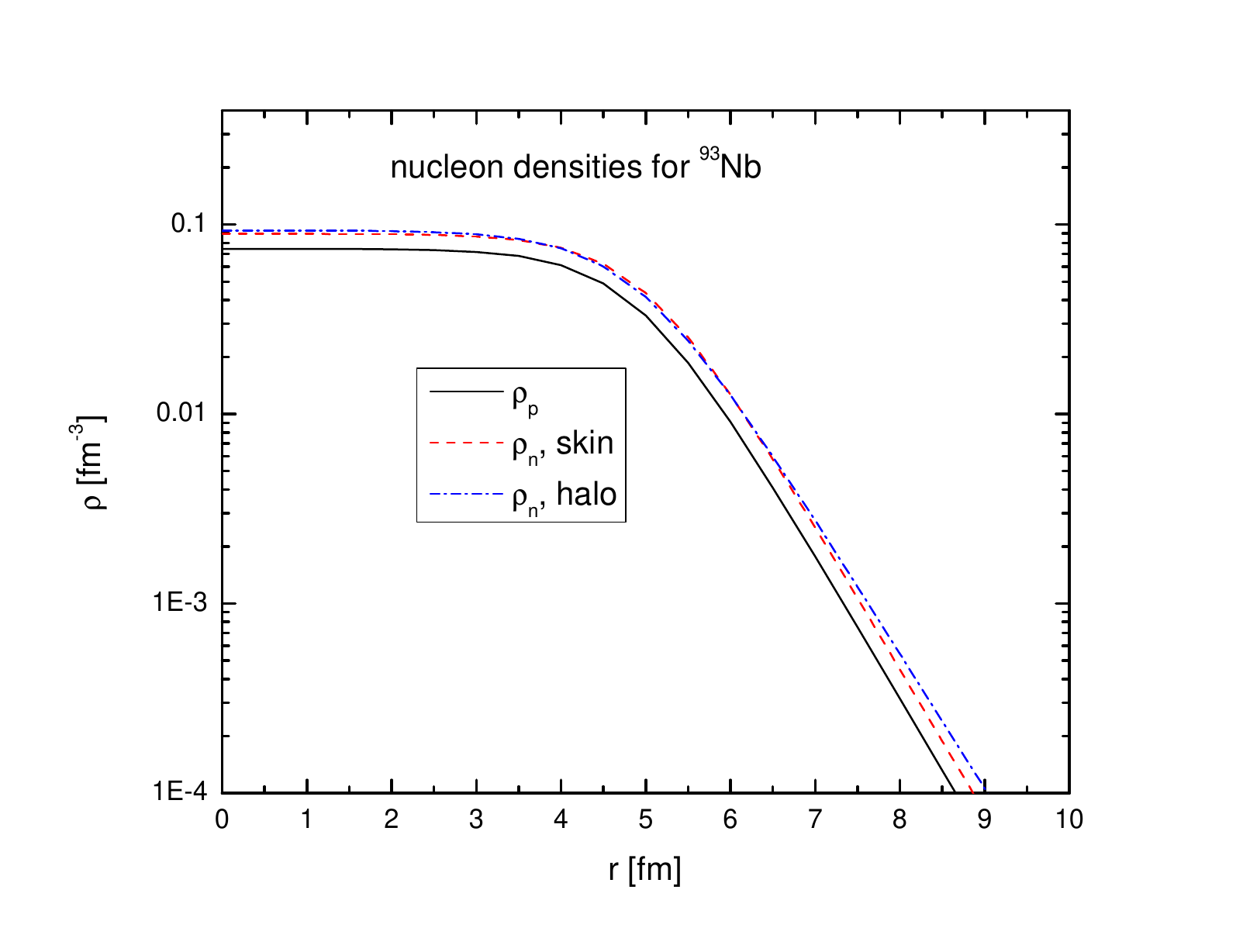}
\vspace*{-2mm} \caption{(Color online.) Proton and neutron densities for $^{93}$Nb. Neutron density
is calculated in the 'skin' and 'halo' forms (see text).}
\label{void}
\end{center}
\end{figure}
\begin{figure}[htb]
\begin{center}
\includegraphics[width=15.0cm]{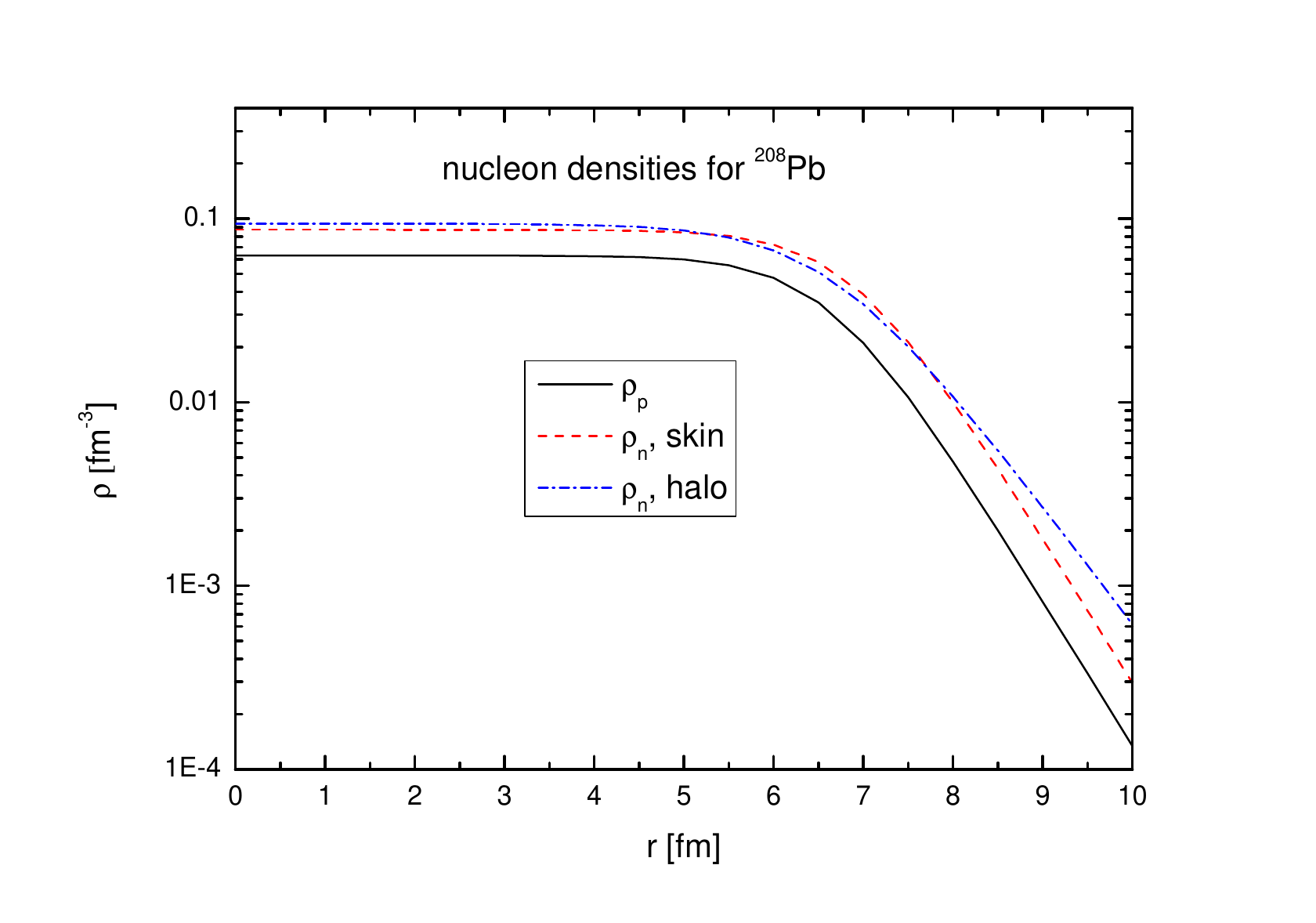}
\vspace*{-2mm} \caption{(Color online.) The same as in Fig. 1, but for the $^{208}$Pb target nucleus.}
\label{void}
\end{center}
\end{figure}
\begin{figure}[htb]
\begin{center}
\includegraphics[width=15.0cm]{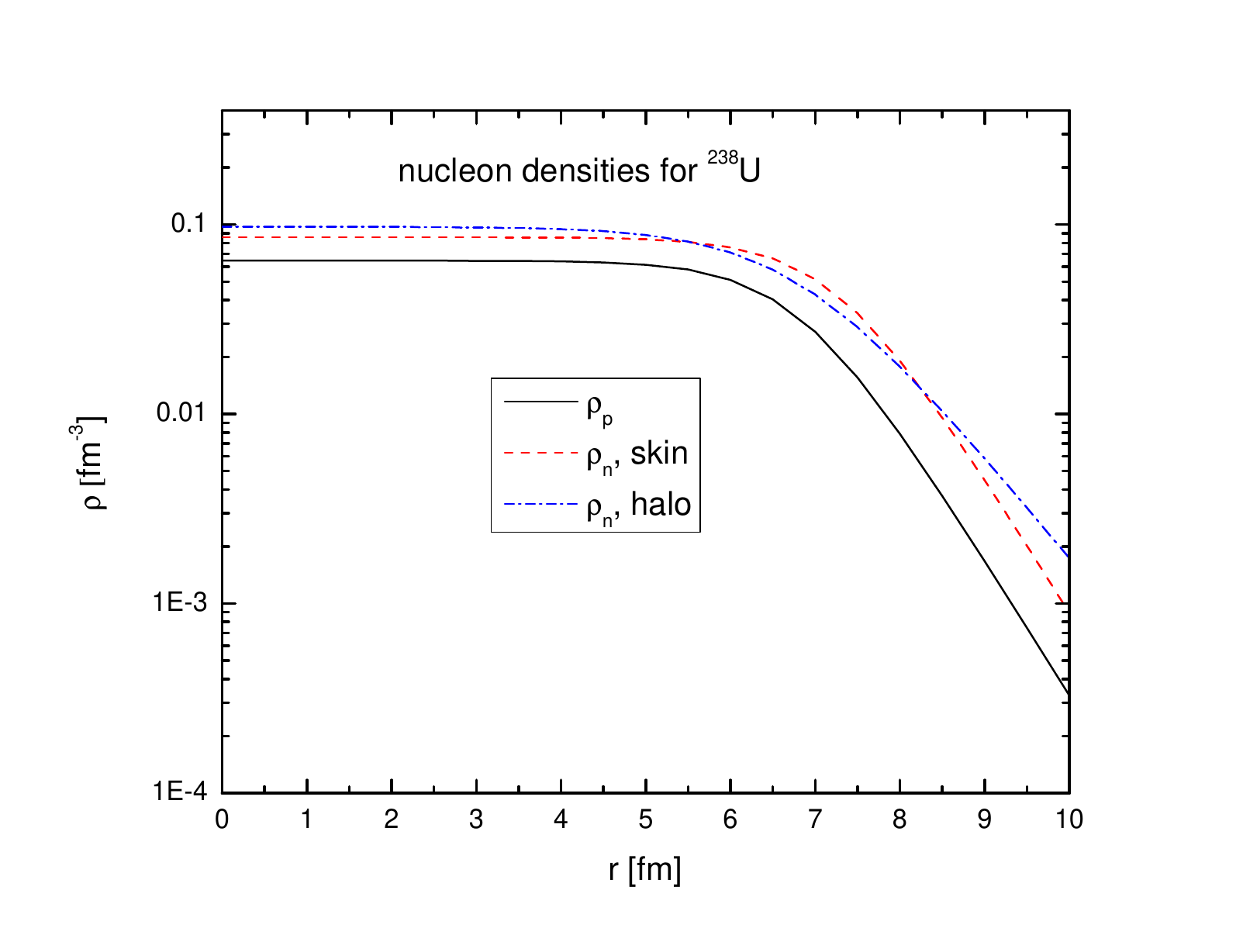}
\vspace*{-2mm} \caption{(Color online.) The same as in Fig. 1, but for the $^{238}$U target nucleus.}
\label{void}
\end{center}
\end{figure}

To see the sensitivity of the results of calculations of the $X(3872)$ meson total production cross sections
on the medium-weight and heavy target nuclei to the choice of the neutron density $\rho_n(r)$, which is not known to sufficient accuracy for these nuclei, we employed for it the 'skin' and 'halo' forms of Ref. [65]. Their radial
parameters $c_n$ and $z_n$ for each nucleus were determined from the r.m.s. radius $r_n$ of $\rho_n(r)$, which assumes in line with Ref. [65] larger value than that $r_p$ for proton density distribution $\rho_p(r)$:
\begin{equation}
r_n-r_p={\gamma}\frac{N-Z}{A}+\delta,
\end{equation}
using the relation [66]
\begin{equation}
c_n=\sqrt{\frac{5}{3}r_n^2-\frac{7}{3}{\pi^2}z_n^2}.
\end{equation}
In Eq. (14) the parameters $\gamma$ and $\delta$ were chosen as $\gamma=1.05$ fm, $\delta=-0.035$ fm and the
r.m.s. radius $r_p$ is considered to be known and was set equal to that of the known nuclear charge density [64].
Then, according to [65], for each value of $r_n$ found from Eq. (14) in the 'skin' form the same diffuseness parameter for protons and neutrons, $z_n=z_p$, is used and the radius parameter $c_n$ is determined from Eq. (15).
In the 'halo' form the same radius parameter, $c_n=c_p$, is assumed and the diffuseness parameter $z_n$
is determined also from Eq. (15). Using the radial parameters for the proton density distributions
given above and their r.m.s. radii: $r_p=4.310$ fm for $^{93}_{41}$Nb,
$r_p=4.655$ fm for $^{112}_{50}$Sn, $r_p=5.420$ fm for $^{184}_{74}$W, $r_p=5.521$ fm for $^{208}_{82}$Pb,
$r_p=5.8434$ fm for $^{238}_{92}$U [64] and going along this way, we obtain the following values for the
radius parameter $c_n$ and for the diffuseness parameter $z_n$, respectively, for the 'skin' and 'halo'
shapes of the neutron distribution for each value of its r.m.s. radius $r_n$: $c_n=4.971$ fm for $^{93}_{41}$Nb,
$c_n=5.487$ fm for $^{112}_{50}$Sn, $c_n=6.745$ fm for $^{184}_{74}$W, $c_n=6.882$ fm for $^{208}_{82}$Pb,
$c_n=7.246$ fm for $^{238}_{92}$U and $z_n=0.6093$ fm for $^{93}_{41}$Nb,
$z_n=0.605$ fm for $^{112}_{50}$Sn, $z_n=0.649$ fm for $^{184}_{74}$W, $z_n=0.673$ fm for $^{208}_{82}$Pb,
$z_n=0.797$ fm for $^{238}_{92}$U. The proton and neutron densities, for example, in $^{93}$Nb, $^{208}$Pb
and $^{238}$U target nuclei, calculated in line with Eq. (13) using two shapes of the neutron distribution
mentioned above, are shown, respectively, in Figs. 1, 2, 3. It is seen that the differences between two
shapes become the most pronounced for all nuclei at distances about 8--10 fm. This will enable us to test
their influence on the observables considered in the present work (see below).

In order to proceed further, we have to specify the $X(3872)$--nucleon absorption cross section $\sigma_{{X(3872)}N}$.
The evaluation of this cross section in different scenarios for the structure of $X(3872)$ proposed in the literature
is a crucial ingredient for its interpretation via, in particular, the $X(3872)$ photoproduction off nuclei since
there is a direct dependence of the rate of this production on the absorption cross section $\sigma_{{X(3872)}N}$
(cf. Eq.(3)), which is expected to be different for different choices for the internal structure of $X(3872)$ (see below). In other words, in our approach the sensitivity to this structure should be encoded in the $X(3872)$ meson production rate. In this exploratory study we consider four popular scenarios for the $X(3872)$ resonance: i) a conventional 2$^3P_1$ $c{\bar c}$ state, the first radial excitation of $\chi_{c1}$, {\it i.e.}, the $\chi_{c1}(2P)$ state [37, 67--78]
\footnote{$^)$It should be mentioned that the bare charmonium state $\chi_{c1}(2P)$ would have the same quantum numbers
$J^{PC}=1^{++}$ as the $X(3872)$, it locates slightly above the $D^0{\bar D}^{*0}$ (${\bar D}^0D^{*0}$) threshold and
has not been found yet [79].}$^)$,
ii) compact, $\sim$ 1 fm, diquark $[cq]$ and antidiquark $[{\bar c}{\bar q}]$ tetraquark state with $q$ either $u$ or $d$ quark [52, 80--90],
iii) a molecular state - a loosely bound, with large spatial size characterized by the r.m.s. radius of the order of 5--10 fm or more,  $S$-wave state mainly in the $D^0{\bar D}^{*0}$+{\it c.c.} system consisting of a neutral $D^0$ and ${\bar D}^{*0}$ or ${\bar D}^0$ and $D^{*0}$ mesons [26--28, 37, 72, 82, 84, 86, 89, 91--105]
\footnote{$^)$In some works (see, for example, Refs. [79, 99, 106--110]), it is assumed that in a molecular picture the $X(3872)$ flavor wave function contains also an additional component $D^+D^{*-}$+{\it c.c.} composed of the charged $D$ and $D^*$ mesons. Because of the proximity of the $X(3872)$ mass to the $D^0{\bar D}^{*0}$ (${\bar D}^0D^{*0}$) threshold,    the $X(3872)$ couples strongly only to the neutral $D^0{\bar D}^{*0}$ and ${\bar D}^0D^{*0}$ channels
(which is manifested by the large branching ratios to the $D^0{\bar D}^{*0}$ and $D^0{\bar D}^{0}\pi^0$ channels
of about 37\% and 40\%, respectively, [111]) and this 'charged' component is of a minor importance in the $X(3872)$ wave function [106--109]. Thus, for instance, in the work [106] it was predicted for the cutoff $\Lambda=0.5$ GeV
that this function contains about 15\% of the $c{\bar c}$ core state, 73\% of the $D^0{\bar D}^{*0}$+{\it c.c.} and 12\% of the $D^+D^{*-}$+{\it c.c.} molecular components.
Therefore, within the molecular hypothesis we will interpret the $X(3872)$ below as a $D^0{\bar D}^{*0}$+{\it c.c.} molecule.}$^)$,
and iv) a hybrid state - a state in which the $c{\bar c}$ charmonium core state $\chi_{c1}(2P)$ couples to the
neutral $D^0{\bar D}^{*0}$ and ${\bar D}^0D^{*0}$ molecular states [37, 73, 106--109, 112--115].

We begin here from a picture of the $X(3872)$ as a pure $c{\bar c}$ state. Since the geometrical size of this state
is expected to be similar to that of the usual charmonium $J/\psi$, whose radius $\sim$ 0.2--0.3 fm (cf. [86, 87, 116] and [117, 118]), and they have the same quark content, it is natural to assume for the $X(3872)$--nucleon absorption cross section $\sigma_{{X(3872)}N}^{{\rm c{\bar c}}}$ in this picture for $X(3872)$ the same absorption cross section as for the charmonium $J/\psi$, {\it i.e.}, $\sigma_{{X(3872)}N}^{{\rm c{\bar c}}}=3.5$ mb [56].

In the second scenario, where the high-momentum (see above) $X(3872)$ resonance is treated as a compact tetraquark state with radius $r_{4q}$, its breakup cross section $\sigma_{{X(3872)}N}^{{\rm 4{q}}}$ in this scenario can be well
approximated by a geometrical cross section $\sigma_{4q}^{\rm geo}={\pi}r_{4q}^2$ [85, 86],
{\it i.e.}, $\sigma_{{X(3872)}N}^{{\rm 4{q}}}\approx{\pi}r_{4q}^2$. For $r_{4q}=0.65$ fm [85, 86],
we obtain that $\sigma_{{X(3872)}N}^{{\rm 4{q}}}=13.3$ mb (cf. [86]).

Within the "pure" molecular interpretation of the $X(3872)$, whose particle content in line with the above-mentioned
is [27, 37, 73, 104, 108]
\begin{equation}
|X(3872)>_{\rm mol}=\frac{1}{\sqrt{2}}\left(|D^0{\bar D}^{*0}>+|{\bar D}^0D^{*0}>\right)
\end{equation}
and in which due to the closeness of the $X(3872)$ to the $D^0{\bar D}^{*0}$ (${\bar D}^0D^{*0}$) threshold
the constituents (charmed mesons) are weakly bound and have large spatial separation in coordinate space
\footnote{$^)$The universal wave function for the $X(3872)$ $D^0{\bar D}^{*0}$ molecule in the coordinate
representation $\psi_{X(3872)}({\bf r})=\exp(-r/a)/(\sqrt{2{\pi}a}r)$ implies that the r.m.s. size $r_{X}$
of this separation is $r_{X}=a/\sqrt{2}$, where the scattering length $a=1/\sqrt{2\mu_0\delta_X}$ can be determined
by knowing the binding energy $\delta_X=m_{D^0}+m_{{\bar D}^{*0}}-m_{X(3872)}$ of the molecule [27, 99, 104].
Here, $\mu_0$ is the $D^0{\bar D}^{*0}$ reduced mass. Taking $\delta_X=0.04\pm0.09$ MeV as input [111, 119], we find that the charmed mesons in the $X(3872)$ have a surprisingly huge r.m.s. separation: $r_{X}=15.9^{+\infty}_{-7.1}$ fm. The typical scale for the relative momentum between the neutral charmed mesons [26, 27, 101],
$\sqrt{2\mu_0\delta_X}\simeq$10--16 MeV/c, is much smaller than their average laboratory momenta $\sim$ 5 GeV/c (see above). With these, we expect that the application (see below) of the quasi-free approximation to estimate the $X(3872)$--nucleon absorption cross section in molecular scenario is sufficiently well justified.}$^)$,
the $X(3872)$--nucleon absorption cross section $\sigma_{{X(3872)}N}^{\rm mol}$ in this interpretation of the $X(3872)$
can be naturally evaluated in the quasi-free approximation [120--122]. In this approximation, the charmed mesons are taken to be on-shell, they are considered to fly together with an average laboratory momentum of the order of 5 GeV/c each
and their binding energy and mutual interactions are neglected. In the quasi-free approximation the $X(3872)$ is absorbed when an intranuclear nucleon (proton or neutron) interacts (elastically or inelastically) with the $D^0$ or with the
${\bar D}^{*0}$ (with the ${\bar D}^0$ or with the $D^{*0}$). In each of these interactions the other charmed meson is a
spectator. Then, assuming that the total cross sections of the free ${\bar D}^{*0}N$ and $D^{*0}N$ interactions are the
same as those for the ${\bar D}^{0}N$ and $D^{0}N$ ones [123--126], we can evaluate the cross section for $X(3872)$ absorption, $\sigma_{{X(3872)}N}^{\rm mol}$, as [123--126]:
\begin{equation}
\sigma_{{X(3872)}N}^{\rm mol}=\sigma_{{X(3872)}p}^{\rm mol}=\sigma_{{X(3872)}n}^{\rm mol},
\end{equation}
where the $X(3872)p$ and $X(3872)n$ absorption cross sections, $\sigma_{{X(3872)}p}^{\rm mol}$ and
$\sigma_{{X(3872)}n}^{\rm mol}$, are given by (cf. [127])
\begin{equation}
\sigma_{{X(3872)}p(n)}^{\rm mol}\approx\sigma_{D^0p(n)}^{\rm el}+\sigma_{D^0p(n)}^{\rm in}+
\sigma_{{\bar D}^0p(n)}^{\rm el}+\sigma_{{\bar D}^0p(n)}^{\rm in}+\sigma_{D^0p \to D^+n}(\sigma_{{\bar D}^0n \to D^-p}).
\end{equation}
Here, $\sigma_{D^0p(n)}^{\rm el(in)}$ and $\sigma_{{\bar D}^0p(n)}^{\rm el(in)}$ are the elastic (inelastic) cross
sections of the free $D^0p$($D^0n$) and ${\bar D}^0p$(${\bar D}^0n$) interactions, respectively. And,
$\sigma_{D^0p \to D^+n}$ and $\sigma_{{\bar D}^0n \to D^-p}$ are the total cross sections of the free charge-exchange
reactions $D^0p \to D^+n$ and ${\bar D}^0n \to D^-p$, correspondingly. In our calculations we adopt for them the
following constants which are relevant to the momentum regime above of 1 GeV/c of interest:
$\sigma_{D^0p(n)}^{\rm el}=\sigma_{{\bar D}^0p(n)}^{\rm el}=10$ mb,
$\sigma_{D^0p(n)}^{\rm in}=10$ mb, $\sigma_{{\bar D}^0p(n)}^{\rm in}=0$,
$\sigma_{D^0p \to D^+n}=\sigma_{{\bar D}^0n \to D^-p}=12$ mb [123--125]. Using these values, we obtain that
$\sigma_{{X(3872)}p}^{\rm mol}=\sigma_{{X(3872)}n}^{\rm mol}=42$ mb and, in view of Eq. (17),
$\sigma_{{X(3872)}N}^{\rm mol}=42$ mb.

In the hybrid approach, it is supposed that the $X(3872)$ wave function in the center of the mass frame is a linear superposition of the charmonim $c{\bar c}$ core state and the $D^0{\bar D}^{*0}$+{\it c.c.} hadronic molecular state
[37, 73, 106, 108, 113]:
\begin{equation}
|X(3872)>_{\rm hyb}=\alpha|c{\bar c}>+\frac{\beta}{\sqrt{2}}\left(|D^0{\bar D}^{*0}>+|{\bar D}^0D^{*0}>\right).
\end{equation}
Here, $\alpha^2$ and $\beta^2$ represent the probabilities to find a charmonium and hadronic configurations,
respectively, for the normalization
\begin{equation}
\alpha^2+\beta^2=1.
\end{equation}
The limiting case of $\alpha^2=0$, $\beta^2=1$ corresponds to the pure molecular interpretation of the $X(3872)$,
while the case of $\alpha^2=1$, $\beta^2=0$ refers to its pure charmonium treatment. In this paper,
following Ref. [37], we will assume that the $X(3872)$ wave function (19) contains 50\% of the genuine nonmolecular $c{\bar c}$ component and 50\% of the molecular $D^0{\bar D}^{*0}$+{\it c.c.} component ($\alpha^2=\beta^2=0.5$).
In addition, to extend the range of applicability of our model and to see the sensitivity of the $X(3872)$
production cross sections from the direct processes (1), (2) to the nonmolecular and molecular probabilities of
the $X(3872)$ in the $c{\bar c}$ and $D^0{\bar D}^{*0}$+{\it c.c.} channels we will further yet adopt in calculations
another additional representative option for them, namely: 15\% and 85\% ($\alpha^2=0.15$ and $\beta^2=0.85$), respectively, as was derived in a model [106]. These two options for the nonmolecular and molecular probabilities of the $X(3872)$ cover the bulk of theoretical and experimental information presently available in this field.
It should be pointed out that it is difficult at present to definitely conclude what are the precise values of these probabilities. The available
presently data definitely rule out the possibility of a dominant nonmolecular component [109]. The precise value of
the molecular probability (which may be as big as 95\% [109]) requires a more accurate determination of the scattering length and effective range of the $D^0{\bar D}^{*0}$ channel, as well as the measurement of these magnitudes for the
$D^+D^{*-}$ channel which have not been determined experimentally up to now [109]. Since the role of the interference
effects between the nonmolecular and molecular contributions to the $X(3872)$--nucleon absorption cross section $\sigma_{{X(3872)}N}^{\rm hyb}$ in the hybrid picture (19) of the $X(3872)$ is expected to be insignificant, we can
represent it (to a good approximation) in the following incoherent probability-weighted sum:
\begin{equation}
\sigma_{{X(3872)}N}^{\rm hyb}=\alpha^2\sigma_{{X(3872)}N}^{\rm c{\bar c}}+\beta^2\sigma_{{X(3872)}N}^{\rm mol}.
\end{equation}
In view of the above, we set $\sigma_{{X(3872)}N}^{\rm c{\bar c}}=3.5$ mb and $\sigma_{{X(3872)}N}^{\rm mol}=42$ mb.
With these values, the $X(3872)$ absorption cross section (21) is $\sigma_{{X(3872)}N}^{\rm hyb}=22.75$ as well as 36.4 mb for the nonmolecular and molecular probabilities of the $X(3872)$ 50\% and 50\% as well as 15\% and 85\%, respectively. We finalize here by summarizing the results obtained above for the $X(3872)$--nucleon absorption cross section $\sigma_{{X(3872)}N}$ in the considered scenarios for $X(3872)$:
\begin{equation}
\sigma_{{X(3872)}N}=\left\{
\begin{array}{lll}
	3.5~{\rm mb}
	&\mbox{for pure $c{\bar c}$ state}, \\
	&\\
    13.3~{\rm mb}
	&\mbox{for tetraquark (4q) state},\\
	&\\
    22.75~{\rm mb}
	&\mbox{for hybrid state (50\%,50\%)},\\
	&\\
    36.4~{\rm mb}
	&\mbox{for hybrid state (15\%,85\%)},\\
	&\\
    42~{\rm mb}
	&\mbox{for $D^0{\bar D}^{*0}$+{\it c.c.} molecule}.
\end{array}
\right.	
\end{equation}
Consequently, the $X(3872)$ as a meson molecule has the largest absorption cross section and, therefore,
is expected to be more easily destroyed than its other states in a hadronic medium.
We will use these values for the quantity $\sigma_{{X(3872)}N}$ throughout our calculations in the
near-threshold energy domain.

 Ignoring the considered nuclear effects: the struck target nucleon binding and Fermi motion, from Eq. (3)
we get the following simple expression for the total cross section
$\sigma_{{\gamma}A\to {X(3872)}X}^{({\rm dir})}(E_{\gamma})$:
\begin{equation}
\sigma_{{\gamma}A\to {X(3872)}X}^{({\rm dir})}(E_{\gamma})=I_{V}[A,\sigma_{{X(3872)}N}]
\sigma_{{\gamma}p \to {X(3872)}p}(\sqrt{s(E_{\gamma})}),
\end{equation}
where the elementary cross section $\sigma_{{\gamma}p \to {X(3872)}p}(\sqrt{s(E_{\gamma})})$ is given above by Eq. (10).
It is valid only at photon energies well above threshold
\footnote{$^)$The difference between the expressions (3) and (23) is quite small (within several percent) at
incident photon energies around 13 GeV and it is $\sim$ 20\% at higher photon energies considered, as our calculations
have shown. On the other hand, it becomes substantial at far subthreshold beam energies
(compare magenta short-dashed and cyan short-dashed-dotted curves in figures 4--7 given below).}$^)$
and allows one to easily estimate here this cross section. As a measure for the
$X(3872)$ absorption cross section $\sigma_{{X(3872)}N}$ in nuclei and, hence, for the $X(3872)$
internal structure we use the so-called $X(3872)$ transparency ratio defined as [128--135]:
\begin{equation}
S_A=\frac{\sigma_{{\gamma}A \to {X(3872)}X}(E_{\gamma})}{A~\sigma_{{\gamma}p \to {X(3872)}p}(\sqrt{s(E_{\gamma})})},
\end{equation}
{\it i.e.} the ratio of the inclusive nuclear $X(3872)$ photoproduction cross section divided by $A$
times the same quantity on a free proton. It should be noticed that this relative observable is more
favorable compared to those based on the absolute cross sections for the aim of getting the information
on the $X(3872)$ nuclear absorption, since it at photon energies well above threshold is mainly sensitive
to the $X(3872)N$ absorption cross section (cf. [130]), whereas the theoretical uncertainties associated
with the $X(3872)$ production mechanism substantially cancel out in it. The direct photon--induced reaction
channels (1), (2) is expected to be dominant in the $X(3872)$ production in ${\gamma}A$ reactions close to threshold
\footnote{$^)$Let's say at incident photon energies $\le$ 13 GeV.
Here, the elementary processes ${\gamma}N \to X(3872)N\pi$ and ${\gamma}N \to X(3872)N{\pi}\pi$
with one and two pions in the final states are expected to be suppressed in $X(3872)$ production in ${\gamma}A$
reactions for kinematics of interest compared to those of (1), (2) due to
their larger production threshold energies ($\approx$12.56 and 13.28 GeV, respectively) in free ${\gamma}N$
interactions. Moreover, since the main inelastic channel in ${\gamma}N$
collisions at these beam energies is the multiplicity production of pions with comparatively
low energies the secondary ${\pi}N \to {X(3872)}X$ processes are energetically suppressed at them as well.
We expect that the full inclusive nuclear $X(3872)$ photoproduction cross section $\sigma_{{\gamma}A \to {X(3872)}X}(E_{\gamma})$, entering into Eq. (24), is entirely exhausted by that of Eq. (3) from the direct $X(3872)$ production channels (1), (2) not only at photon energies $E_{\gamma} \le 13$ GeV, but also for higher beam energies considered in the present work.}$^)$.
Then, according to Eqs. (3) and (4), we have:
\begin{equation}
S_A=\frac{\sigma_{{\gamma}A \to {X(3872)}X}^{({\rm dir})}(E_{\gamma})}{A~\sigma_{{\gamma}p \to {X(3872)}p}(\sqrt{s(E_{\gamma})})}=
\frac{I_{V}[A,\sigma_{{X(3872)}N}]}{A}\frac{
\left<\sigma_{{\gamma}p \to {X(3872)}p}(E_{\gamma})\right>_A}{\sigma_{{\gamma}p \to {X(3872)}p}(\sqrt{s(E_{\gamma})})}.
\end{equation}
At photon energies around 13 GeV the difference between the cross sections $\left<\sigma_{{\gamma}p \to {X(3872)}p}(E_{\gamma})\right>_A$ and $\sigma_{{\gamma}p \to {X(3872)}p}(\sqrt{s(E_{\gamma})})$ can be ignored (cf. footnote 10)).
Then, from Eq. (25) we obtain (in a good approximation):
\begin{equation}
S_A=\frac{1}{A}I_{V}[A,\sigma_{{X(3872)}N}].
\end{equation}
It is usually customary to normalize the transparency ratio of the meson to a light nucleus like $^{12}$C
[135--140].
With this, for $X(3872)$ we have:
\begin{equation}
T_A=\frac{S_A}{S_{\rm C}}=\frac{12~\sigma_{{\gamma}A \to {X(3872)}X}(E_{\gamma})}{A~\sigma_{{\gamma}{\rm C} \to {X(3872)}X}(E_{\gamma})}.
\end{equation}
Here, $\sigma_{{\gamma}A \to {X(3872)}X}(E_{\gamma})$ and $\sigma_{{\gamma}{\rm C} \to {X(3872)}X}(E_{\gamma})$
are the inclusive total cross sections for $X(3872)$ production in
${\gamma}A$ and ${\gamma}{\rm C}$ collisions at incident photon energy $E_{\gamma}$, respectively.
In accordance with the above, the expression (27) for the normalized to carbon transparency ratio $T_A$ can be
represented in the following form:
\begin{equation}
T_A=\frac{12~\sigma_{{\gamma}A \to {X(3872)}X}^{({\rm dir})}(E_{\gamma})}{A~\sigma_{{\gamma}{\rm C} \to {X(3872)}X}^{({\rm dir})}(E_{\gamma})}=
\frac{12~I_V[A,\sigma_{{X(3872)}N}]}{A~I_V[{\rm C},\sigma_{{X(3872)}N}]}
\frac{\left<\sigma_{{\gamma}p \to {X(3872)}p}(E_{\gamma})\right>_A}
{\left<\sigma_{{\gamma}p \to {X(3872)}p}(E_{\gamma})\right>_{\rm C}}.
\end{equation}
Ignoring the difference between the cross sections $\left<\sigma_{{\gamma}p \to {X(3872)}p}(E_{\gamma})\right>_A$
and $\left<\sigma_{{\gamma}p \to {X(3872)}p}(E_{\gamma})\right>_{\rm C}$, from (28) we approximately get
\footnote{$^)$It is easily seen that, according to the equations (23) and (28), the expression (29) describes the
transparency ratio $T_A$ also in the case when direct ${\gamma}N \to X(3872)N$ processes proceed on a free target
nucleon being at rest.}$^)$
:
\begin{equation}
T_A \approx \frac{12~I_V[A,\sigma_{{X(3872)}N}]}{A~I_V[{\rm C},\sigma_{{X(3872)}N}]}.
\end{equation}
As is easy to see, the integral (4) for the quantity $I_V[A,\sigma_{{X(3872)}N}]$ can be transformed to a simpler expression:
\begin{equation}
I_{V}[A,\sigma_{{X(3872)}N}]=\frac{\pi}{\sigma_{{X(3872)}N}}\int\limits_{0}^{R^2}dr_{\bot}^2
\left(1-e^{-\sigma_{{X(3872)}N}\int\limits_{-\sqrt{R^2-r_{\bot}^2}}^{\sqrt{R^2-r_{\bot}^2}}
\rho(\sqrt{r_{\bot}^2+x^2})dx}\right),
\end{equation}
which in the case of a uniform nucleon densities for a nucleus of a radius $R=r_0A^{1/3}$ with a sharp boundary is reduced to even more simple form:
\begin{equation}
I_V[A,\sigma_{{X(3872)}N}]=\frac{3A}{2a_1}\left\{1-\frac{2}{a_1^2}[1-(1+a_1)e^{-a_1}]\right\}, \,\,\,\,a_1=3A\sigma_{X(3872)N}/2{\pi}R^2.
\end{equation}
The simple formulas (23) and (31) allow one to easily estimate the total cross section $\sigma_{{\gamma}A\to {X(3872)}X}^{({\rm dir})}(E_{\gamma})$ at above threshold energies.

Before closing this subsection, we discuss now the momentum-dependent differential cross section for $X(3872)$
production from the direct processes (1) and (2) in ${\gamma}A$ reactions.
At the incident photon energies of interest the $X(3872)$ mesons are produced at very small
laboratory polar angles
\footnote{$^)$Thus, for example, the maximum angle of $X(3872)$ meson production off a free proton at rest in reaction (1) is about 3.4$^{\circ}$ at photon energy of 13 GeV.}$^)$
. Therefore, we will calculate their momentum distribution from considered target nuclei
for the laboratory solid angle ${\Delta}{\bf \Omega}_{X(3872)}$ = $0^{\circ} \le \theta_{X(3872)} \le 10^{\circ}$,
and $0 \le \varphi_{X(3872)} \le 2{\pi}$. Then, accounting for the results presented both in Ref. [141] and above by
Eqs. (3)--(8), we can get the following expression for this distribution:
\begin{equation}
\frac{d\sigma_{{\gamma}A\to {X(3872)}X}^{({\rm dir})}
(p_{\gamma},p_{X(3872)})}{dp_{X(3872)}}=
2{\pi}I_{V}[A,\sigma_{{X(3872)}N}]
\int\limits_{\cos10^{\circ}}^{1}d\cos{{\theta_{X(3872)}}}
\left<\frac{d\sigma_{{\gamma}p\to {X(3872)}{p}}(p_{\gamma},
p_{X(3872)},\theta_{X(3872)})}{dp_{X(3872)}d{\bf \Omega}_{X(3872)}}\right>_A,
\end{equation}
where
$\left<\frac{d\sigma_{{\gamma}p \to {X(3872)}p}(p_{\gamma},
p_{X(3872)},\theta_{X(3872)})}{dp_{X(3872)}d{\bf \Omega}_{X(3872)}}\right>_A$
is the off-shell inclusive differential cross section for the production of $X(3872)$ mesons
with momentum ${\bf p}_{X(3872)}$ in the channel ${\gamma}p \to {X(3872)}p$,
averaged over the Fermi motion and binding energy of the protons in the nucleus.
It can be expressed by Eqs. (28), (31)--(39) from Ref. [141], in which one needs to make the
substitution: $\Upsilon(1S) \to X(3872)$. For better readability of this paper, we do not give
these expressions here. In order to calculate the c.m. $X(3872)$ angular distribution in process (1)
(cf. Eq. (34) from Ref. [141]) one should know its exponential $t$-slope parameter $b_{X(3872)}$ in
the near-threshold energy region. Since the $X(3872)$ angular distribution from the reaction
${\gamma}p \to {X(3872)}p$ is experimentally unknown and because of the similarity of
the $X(3872)$ production rate with the production rate of the another radially excited $\psi(2S)$ meson in hadronic
collisions (see above), we will assume that the slope parameter $b_{X(3872)}$ is the same as an
exponential $t$-slope $b_{\psi(2S)}$ of the differential cross section of the reaction ${\gamma}p \to {\psi(2S)}p$
in the c.m. system near the threshold
\footnote{$^)$The threshold photon energy for ${\psi(2S)}p$ production on a free proton being at rest is 10.93 GeV.}$^)$
.
The slope for 2S-radially excited heavy charmonium elastic photoproduction was found [142, 143] to have a smaller value than for $J/\psi$:
\begin{equation}
b_{\psi(2S)}(s(E_{\gamma}))=b_{J/\psi}(s(E_{\gamma}))-\Delta_b(s(E_{\gamma})).
\end{equation}
For the c.m. free space energy behavior of the diffraction slope $b_{J/\psi}(s(E_{\gamma}))$ we use the standard Regge form [142, 143]:
\begin{equation}
b_{J/\psi}(s(E_{\gamma}))=b_0+2{\alpha}^{\prime}(0){\rm ln}\left(\frac{s(E_{\gamma})}{s_0}\right),
\end{equation}
where the parameters ${\alpha}^{\prime}(0)=0.171~{\rm GeV}^{-2}$ and $b_0=1.54~{\rm GeV}^{-2}$ were fitted in [142]
to high-energy data on $J/\psi$ photoproduction with $s_0=1$~GeV$^2$.
It is worth noting that the extrapolation of the simple fit (34) of the high-energy data to the ${J/\psi}p$ threshold energies (to the energies around 10 GeV) is also compatible with available here data for an exponential $J/\psi$ $t$-slope $b_{J/\psi}$ [141]. In line with Eq. (34), for $E_{\gamma}=13$ GeV we have that
$b_{J/\psi} \approx 2.7$ GeV$^{-2}$. The factor $\Delta_b(s(E_{\gamma}))$, entering into the Eq. (33), was parameterized
in Ref. [142] as:
\begin{equation}
\Delta_b^T(s(E_{\gamma}))=0.60-0.04{\rm ln}\left(\frac{s(E_{\gamma})}{s_0}\right)~{\rm GeV}^{-2},\,\,\,\,\,\,
\Delta_b^L(s(E_{\gamma}))=1.53-0.12{\rm ln}\left(\frac{s(E_{\gamma})}{s_0}\right)~{\rm GeV}^{-2}
\end{equation}
for photoproduction of $T$ and $L$ polarized $\psi(2S)$ mesons. Numerically, $\Delta_b^{T/L}$ are obtained as:
\begin{equation}
\Delta_b^T=0.47~{\rm GeV}^{-2},\,\,\,\,\,\,\,\,
\Delta_b^L=1.14~{\rm GeV}^{-2}
\end{equation}
for photon energy $E_{\gamma}=13$ GeV.
A weighted average of the two values (36) is about 0.7 GeV$^{-2}$, {\it i.e.}, the $t$-slope parameter
$b_{\psi(2S)}\approx2.0$ GeV$^{-2}$ for incident photon beam energy of 13 GeV. And, thus, in line with the aforementioned, the value of the $X(3872)$ slope parameter $b_{X(3872)}$ is $b_{X(3872)}\approx2.0$ GeV$^{-2}$
at this energy. We will employ it in our subsequent differential cross-section calculations.
\begin{figure}[!h]
\begin{center}
\includegraphics[width=15.0cm]{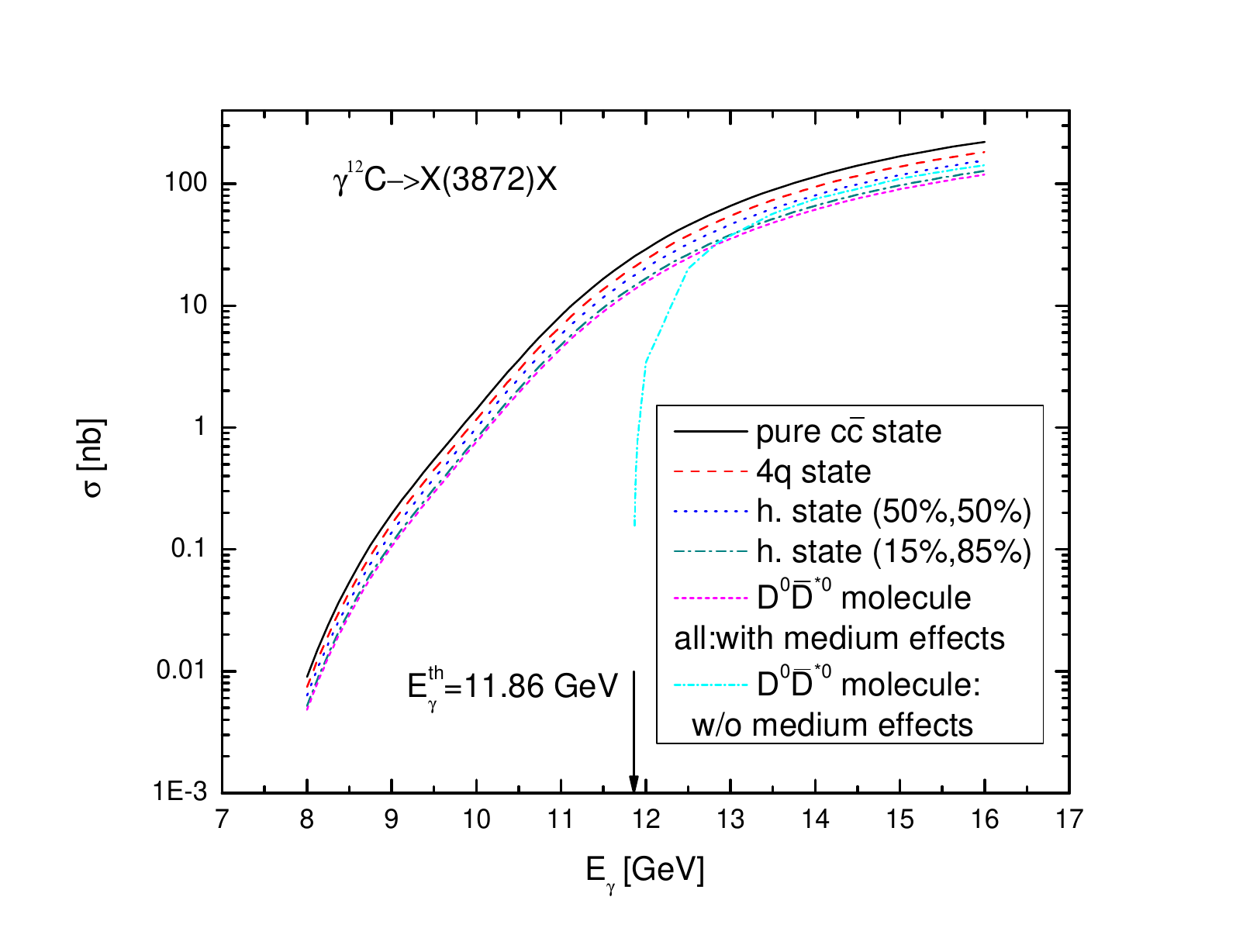}
\vspace*{-2mm} \caption{(Color online.) Excitation function for production of $X(3872)$
mesons off $^{12}$C from primary ${\gamma}N \to {X(3872)}N$ reactions proceeding on an
off-shell target nucleons and on a free ones being at rest.
The curves are calculations in the scenarios, in which the $X(3872)$ is treated as a purely $c{\bar c}$ charmonium state, as a tetraquark state: a compact four quark (4$q$) state, as a purely molecular state: a weakly coupled
$D^0{\bar D}^{*0}$ molecule, or as a hybrid state: a mixture of the $c{\bar c}$ and $D^0{\bar D}^{*0}$ states
in which there are, respectively, 50\% of the $c{\bar c}$ component and 50\% molecular component as well as
15\% and 85\% of the nonmolecular and molecular states.
The arrow indicates the threshold energy for $X(3872)$ photoproduction on a free nucleon.}
\label{void}
\end{center}
\end{figure}
\begin{figure}[!h]
\begin{center}
\includegraphics[width=15.0cm]{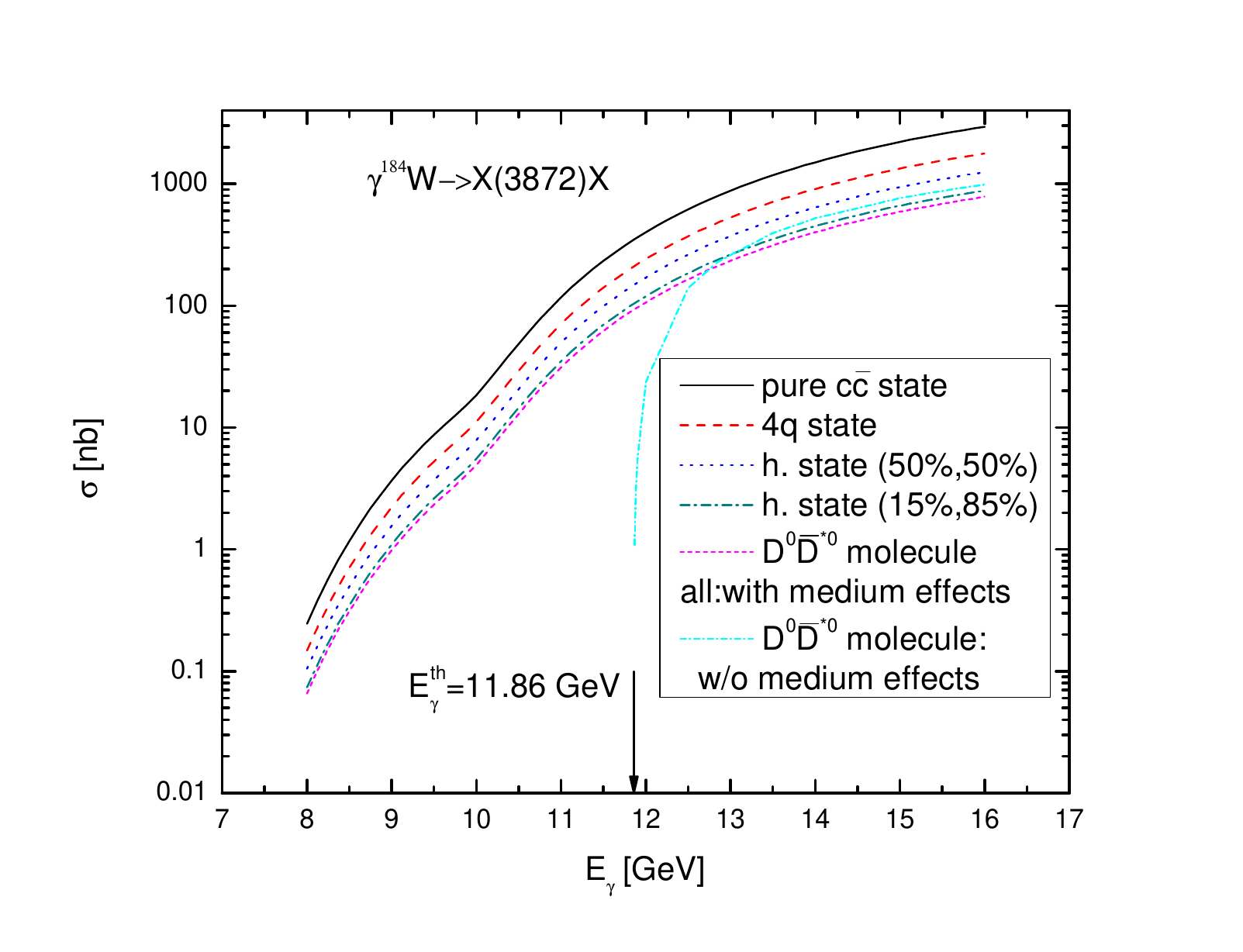}
\vspace*{-2mm} \caption{(Color online.) The same as in Fig. 4, but for the $^{184}$W target nucleus.}
\label{void}
\end{center}
\end{figure}
\begin{figure}[!h]
\begin{center}
\includegraphics[width=15.0cm]{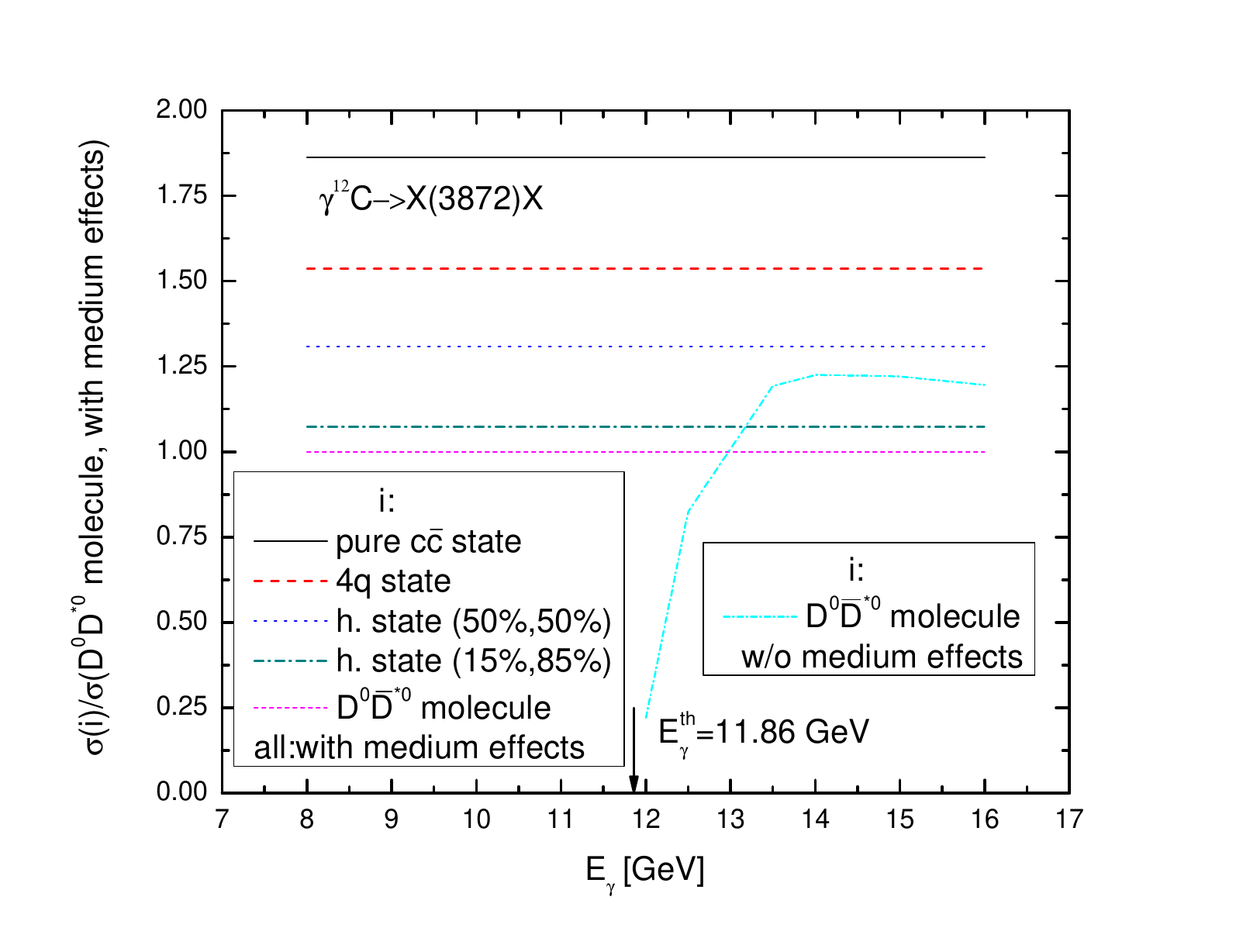}
\vspace*{-2mm} \caption{(Color online.) Ratio between the $X(3872)$ production cross sections on $^{12}$C,
shown in Fig. 4, and the cross section, calculated in the molecular scenario of $X(3872)$ for
an off-shell target nucleons, as a function of photon energy. The arrow indicates the threshold for the reaction ${\gamma}N \to {X(3872)}N$ proceeding on a free target nucleon.}
\label{void}
\end{center}
\end{figure}
\begin{figure}[!h]
\begin{center}
\includegraphics[width=15.0cm]{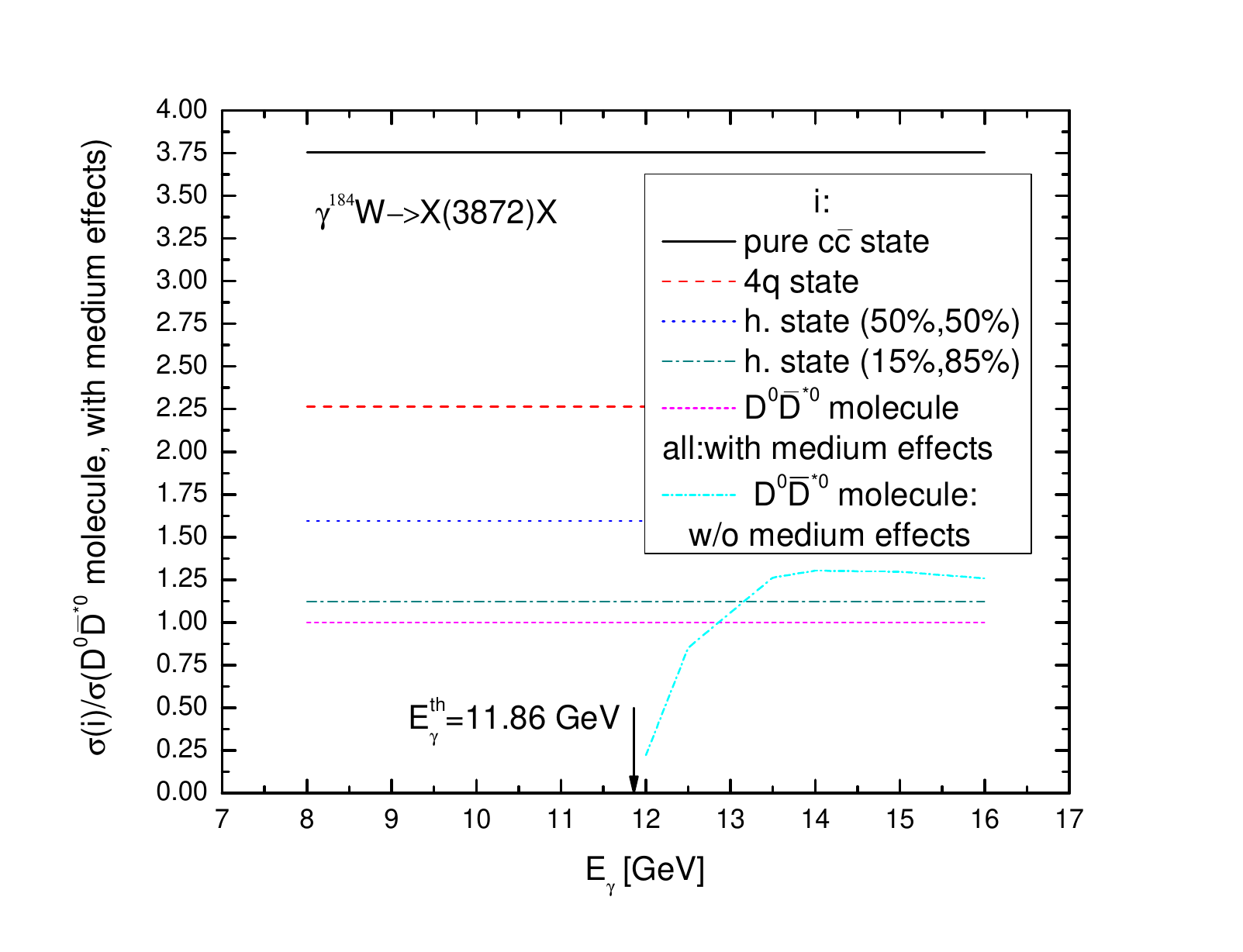}
\vspace*{-2mm} \caption{(Color online.) Ratio between the $X(3872)$ production cross sections on $^{184}$W,
shown in Fig. 5, and the cross section, calculated in the molecular scenario of $X(3872)$ for
an off-shell target nucleons, as a function of photon energy. The arrow indicates the threshold for the reaction ${\gamma}N \to {X(3872)}N$ proceeding on a free target nucleon.}
\label{void}
\end{center}
\end{figure}
\begin{figure}[!h]
\begin{center}
\includegraphics[width=15.0cm]{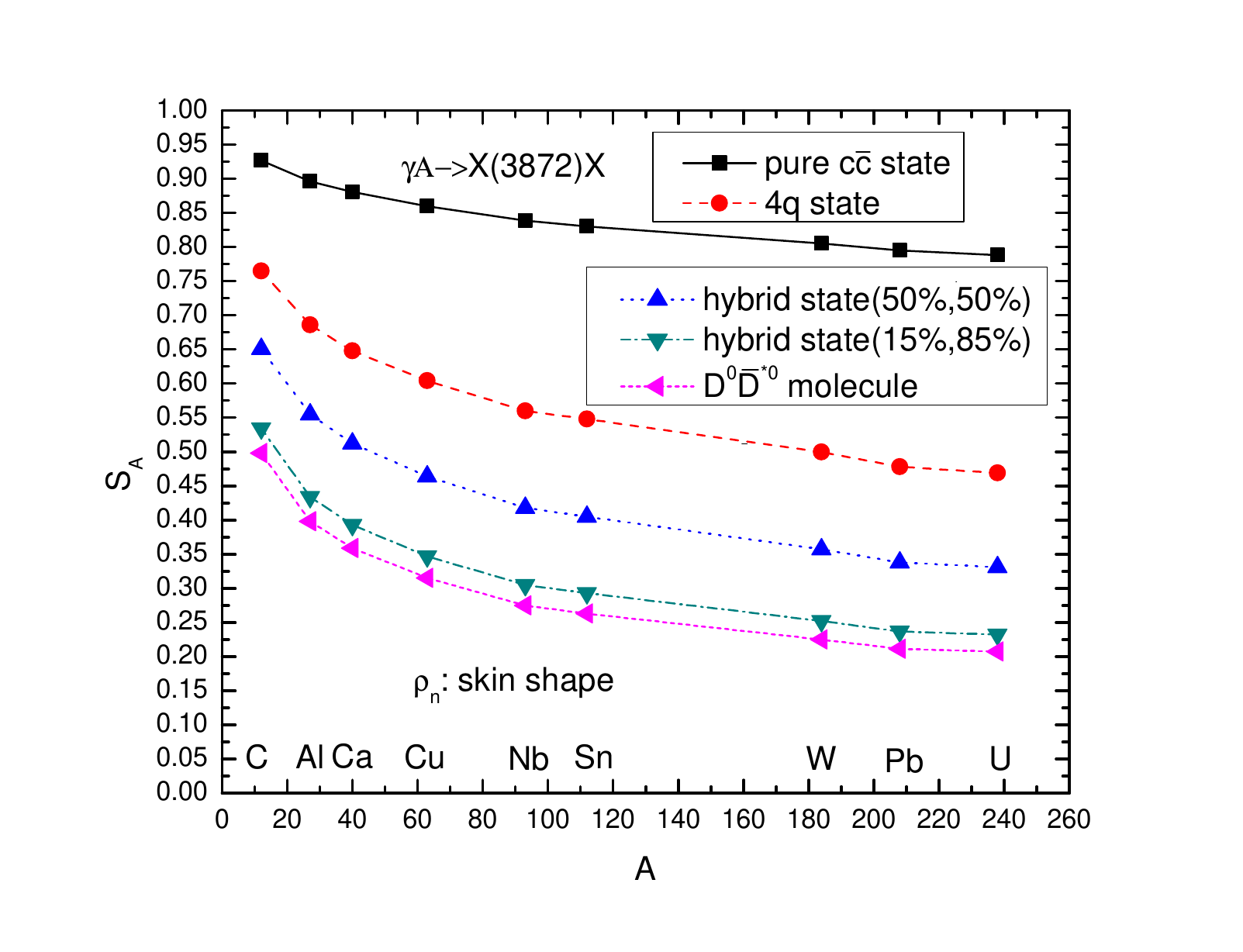}
\vspace*{-2mm} \caption{(Color online.) Transparency ratio $S_A$ for the $X(3872)$ mesons
from primary ${\gamma}N \to {X(3872)}N$ reactions proceeding on a free target nucleons being at rest
as a function of the nuclear mass number $A$ in the considered scenarios for the $X(3872)$ internal structure.
The lines are to guide the eyes.}
\label{void}
\end{center}
\end{figure}
\begin{figure}[!h]
\begin{center}
\includegraphics[width=15.0cm]{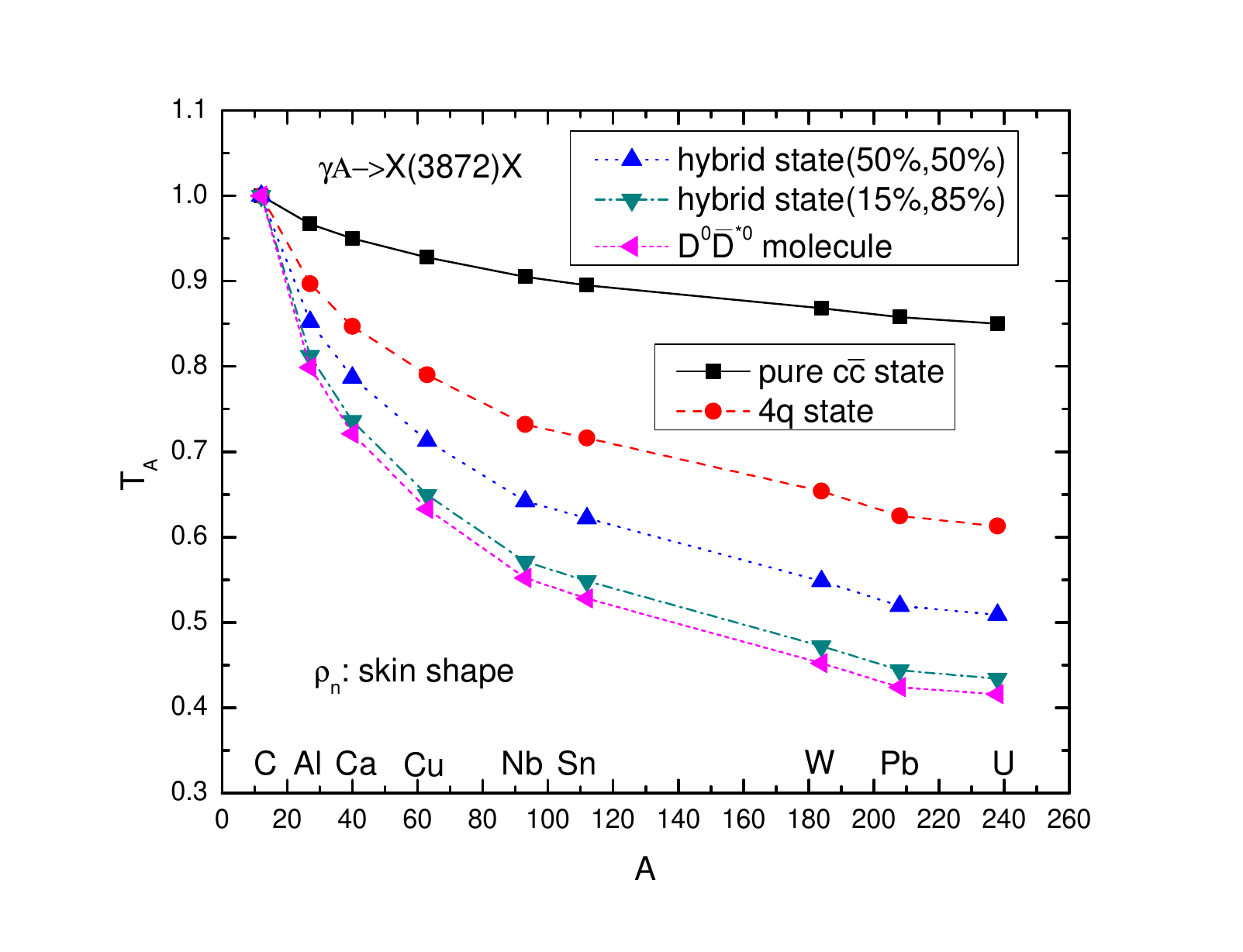}
\vspace*{-2mm} \caption{(Color online.) Transparency ratio $T_A$ for the $X(3872)$ mesons
from primary ${\gamma}N \to {X(3872)}N$ reactions proceeding on a free target nucleons being at rest
as a function of the nuclear mass number $A$ in the considered scenarios for the $X(3872)$ internal structure.
The lines are to guide the eyes.}
\label{void}
\end{center}
\end{figure}
\begin{figure}[!h]
\begin{center}
\includegraphics[width=15.0cm]{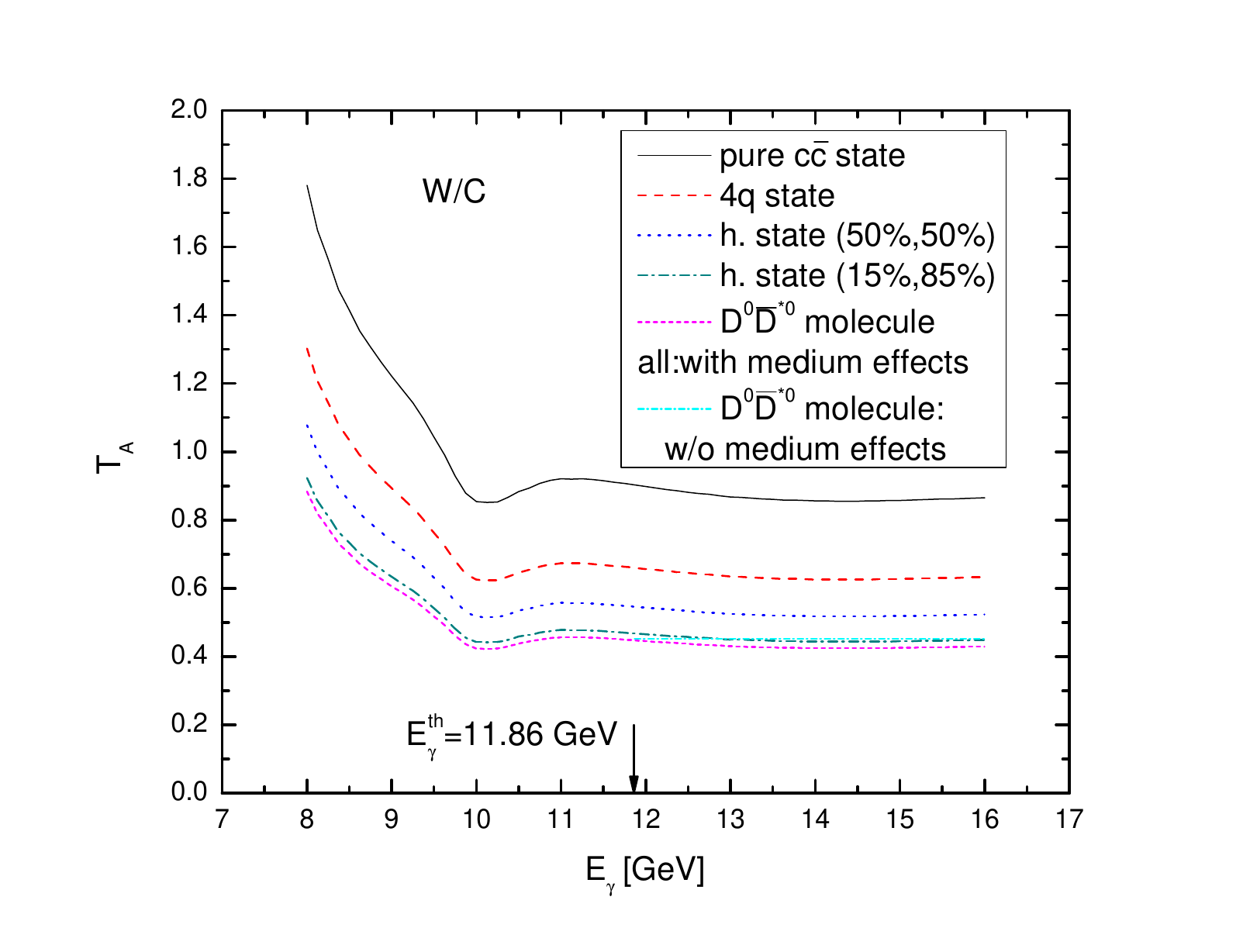}
\vspace*{-2mm} \caption{(Color online.) Transparency ratio $T_A$ for the $X(3872)$ mesons
from primary ${\gamma}N \to {X(3872)}N$ reactions proceeding on an off-shell and free target nucleons
as a function of the incident photon energy for combination $^{184}$W/$^{12}$C
in the considered scenarios for the $X(3872)$ internal structure. The arrow indicates the threshold energy
for $X(3872)$ photoproduction on a free target nucleon at rest.}
\label{void}
\end{center}
\end{figure}
\begin{figure}[!h]
\begin{center}
\includegraphics[width=15.0cm]{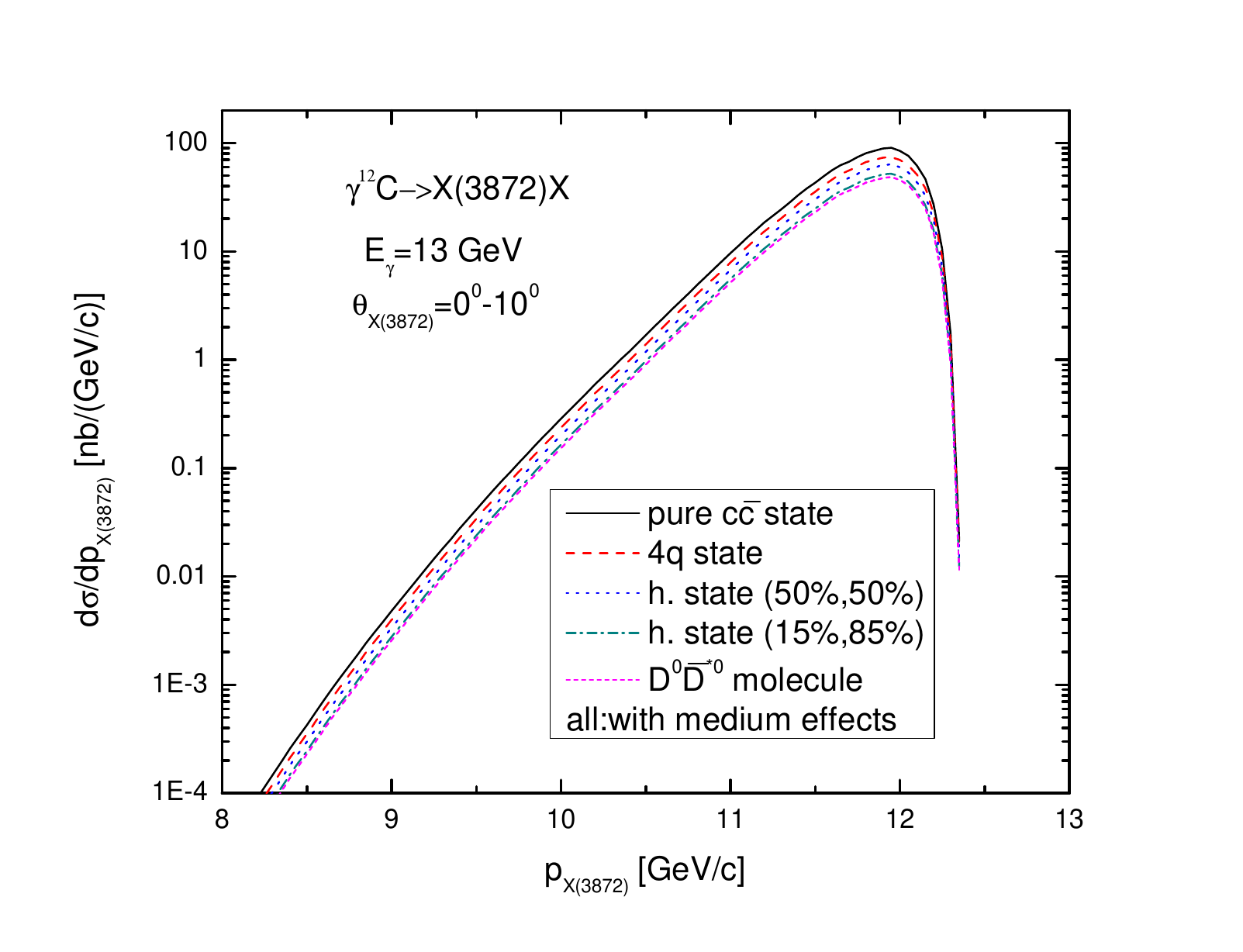}
\vspace*{-2mm} \caption{(Color online.) The direct momentum distribution of $X(3872)$ mesons,
produced in the reaction $\gamma$$^{12}$C $\to X(3872)X$ in the laboratory polar angular range of
0$^{\circ}$--10$^{\circ}$ and calculated in line with Eq. (32) at initial photon energy of 13 GeV in the
laboratory system. The curves are calculations in the scenarios, in which the $X(3872)$ is treated as a purely $c{\bar c}$ charmonium state, as a tetraquark state: a compact four quark (4$q$) state, as a purely molecular state: a weakly coupled $D^0{\bar D}^{*0}$ molecule, or as a hybrid state: a mixture of the $c{\bar c}$ and $D^0{\bar D}^{*0}$ states
in which there are, respectively, 50\% of the $c{\bar c}$ component and 50\% molecular component as well as
15\% and 85\% of the nonmolecular and molecular states.}
\label{void}
\end{center}
\end{figure}
\begin{figure}[!h]
\begin{center}
\includegraphics[width=15.0cm]{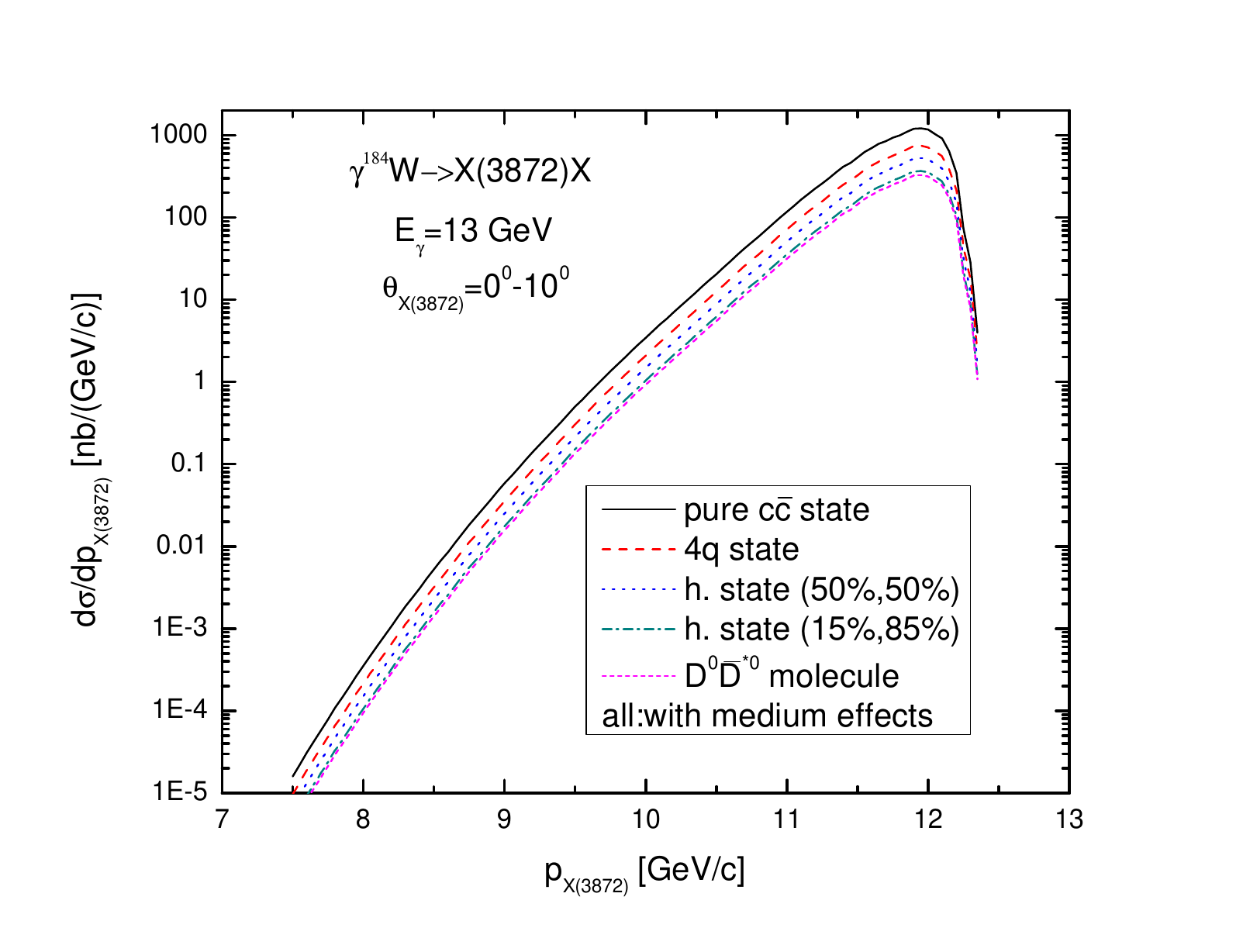}
\vspace*{-2mm} \caption{(Color online.) The same as in Fig. 11, but for the $^{184}$W target nucleus.}
\label{void}
\end{center}
\end{figure}

\section*{3. Results and discussion}

\hspace{1.5cm}At first, we consider the excitation functions for production of $X(3872)$ mesons off
$^{12}$C and $^{184}$W target nuclei. They were calculated on the basis of equation (3)
for five basic adopted options for the $X(3872)$ absorption cross section $\sigma_{{X(3872)}N}$ in nuclei
(cf. Eq. (22)) as well as in line with formula (23) for the free target nucleon
being at rest with the value of $\sigma_{X(3872) N}=42$ mb corresponding to the molecular scenario for $X(3872)$,
and are given in figures 4 and 5, respectively. For heavy tungsten nucleus for the neutron density $\rho_n(r)$
the 'skin' form of Ref. [65] (see also above) was used in the calculations
\footnote{$^)$These calculations show that the replacement of the 'skin' form of the neutron density in the case of the $^{184}$W target nucleus by the 'halo' one made little difference: the calculated $X(3872)$ total production cross
sections on this nucleus are increased only by a small fractions about 1--2\%. An analogous small fractions are available in the cases of another medium-mass and heavy target nuclei considered in the present work. Therefore, we will further report for $^{184}$W and these nuclei the numerical results obtained only for the 'skin' form of their neutron densities.}$^)$
.
It is seen that the difference between calculations with and without
accounting for the target nucleon Fermi motion (between magenta short-dashed and cyan short-dashed-dotted curves)
is indeed sufficiently small at above threshold photon energies $\sim$ 12.5--16 GeV and especially at energies around
13 GeV, while at lower incident energies its influence on $X(3872)$ yield is quite essential.
It is seen yet from these figures that the $X(3872)$ meson total cross sections reveal a certain sensitivity to the
$X(3872)$--nucleon absorption cross section $\sigma_{{X(3872)}N}$ and, hence, to the $X(3872)$
inner structure, for both target nuclei and for photon energies both below and above threshold energy of 11.86 GeV.
The absolute values of the total cross sections have at above threshold photon energies of $\sim$ 12.5--16 GeV
a well measurable strengths $\sim$ 30--200 and 200--3000 nb for carbon and tungsten target nuclei, correspondingly.
Here there are a well separated and experimentally distinguishable
differences (especially for the heavy target nucleus $^{184}$W) between the considered choices:
$\sigma_{{X(3872)}N}=3.5$ mb and $\sigma_{{X(3872)}N}=13.3$ mb, $\sigma_{{X(3872)}N}=13.3$ mb and $\sigma_{{X(3872)}N}=22.75$ mb, $\sigma_{{X(3872)}N}=22.75$ mb and $\sigma_{{X(3872)}N}=36.4$ mb, $\sigma_{{X(3872)}N}=36.4$ mb and $\sigma_{{X(3872)}N}=42$ mb or between the considered scenarios:
pure $c{\bar c}$ state and tetraquark state, tetraquark state and hybrid state with the nonmolecular
and molecular probabilities of 50\% and 50\%,
hybrid state with the nonmolecular and molecular probabilities of 50\% and 50\% and
hybrid state with the nonmolecular and molecular probabilities of 15\% and 85\%,
hybrid state with the nonmolecular and molecular probabilities of 15\% and 85\% and pure molecular state
for the $X(3872)$--nucleon absorption cross section or for the $X(3872)$ internal structure.
They are $\sim$ 21, 17, 22, 8\% and 66, 42, 42, 12\% in the cases of $^{12}$C  and $^{184}$W target nuclei,
respectively. Therefore, one might expect to measure both these strengths and these differences
in the future high-precision experiments at the CEBAF facility.
To further motivate the conducting of such experiments at this
facility, it is highly desirable to evaluate the $X(3872)$ production rates in the ${\gamma}^{12}$C and
${\gamma}^{184}$W reactions. For this purpose, we translate the $X(3872)$ photoproduction total cross section
predictions, reported above, into the expected yields of the $X(3872)$ signals from the reactions
${\gamma}^{12}{\rm C}(^{184}{\rm W}) \to X(3872)X$, $X(3872) \to {J/\psi}\pi^+\pi^-$, $J/\psi \to l^+l^-$
\footnote{$^)$The lepton pairs $l^+l^-$ denote both $\mu^+\mu^-$ and $e^+e^-$.}$^)$
.
We assume that a conservative integrated luminosity of about 100 pb$^{-1}$ could be reached for a year of data taking
\footnote{$^)$Thus, for example, one year of running of the GlueX (2023) experiment [60] devoted to the study
of the ${\gamma}p \to {J/\psi}p$ reaction in the full near-threshold kinematic region at the CEBAF facility
delivered a total accumulated luminosity of about 320 pb$^{-1}$.}$^)$
.
For the $X(3872)$ yield (for the total number of the $X(3872)$ events) estimates in a one-year run, one needs
to multiply the above luminosity by the $X(3872)$ production total cross sections of 30--200 and 200--3000 nb
on the carbon and tungsten target nuclei, respectively, as well as by the detection efficiency and by the
appropriate branching ratios $Br[X(3872) \to {J/\psi}\pi^+\pi^-]\approx$4.1\% [90] and
$Br[{J/\psi} \to \l^+l^-]\approx$12\%. Even with a relatively low (and realistic) 10\% detection efficiency,
we estimate about of 1476--9840 and 9840--147600 events per year for the $X(3872)$ signal in the cases of the $^{12}$C and $^{184}$W target nuclei, respectively. We see that a considerable amount of $X(3872)$ could be observed
\footnote{$^)$Since the $X(3872)$ photoproduction total cross sections of the
elementary reaction ${\gamma}p \to X(3872)p$ have an absolute values $\sim$ 10--20 nb at photon energies well
above threshold of interest [47, 58], the production rate of the $X(3872)$ in this reaction is expected to be
also large. Accounting for these $X(3872)$ photoproduction total cross sections and the estimates of the $X(3872)$
yields given above, we can evaluate the numbers of events expected in the measurements of the $X(3872)$ mesons
in the considered energy range for this reaction as about of 500--1000 with the one-year run. These numbers are commensurate with those of the $J/\psi$ charmonium, measured in the reaction ${\gamma}p \to {J/\psi}p$
in the GlueX(2023) experiment [60] at JLab. Thus, the values of the total cross section of this reaction, measured
in this experiment, are about of 0.1--1 nb near the threshold. Using these values as well as values of the integrated
luminosity of 320 pb$^{-1}$, of the detection efficiency of 10\% and of the branching ratio $Br[J/\psi \to e^+e^-]$ of 6\%, we can estimate the measured numbers of the $J/\psi$ events in this experiment as 180--1800 (cf. [60]). Therefore, it will be possible to measure the $X(3872)$ state even through the scan of its total photoproduction cross section on a proton target in the near-threshold energy region, at least, in the future CEBAF experiments.}$^)$
.
Hence, the measurements of the photonuclear $X(3872)$ meson production might help to decipher its
internal structure.

To see more clearly the sensitivity of the total cross sections, presented in Figs. 4 and 5, to the
$X(3872)$--nucleon absorption cross section, we show in Figs. 6 and 7 the energy dependences of the
ratios of these cross sections to the cross section, calculated in the molecular scenario for $X(3872)$
for an off-shell target nucleons, on a linear scale for $^{12}$C and $^{184}$W target nuclei, respectively.
It is nicely seen from these figures that there are indeed experimentally distinguishable differences
between the results corresponding to the considered options for the $X(3872)$ meson--nucleon absorption
cross section or to the considered configurations for its inner structure practically for both target nuclei
and for both subthreshold and above threshold incident photon energies. This means that the structure of $X(3872)$
mesons could be investigated at the CEBAF through the energy dependence of their absolute (and relative)
production cross sections in inclusive ${\gamma}A$ reactions in the near-threshold energy region.
Figures 6 and 7 also show convincingly that the difference between the absolute $X(3872)$ production cross sections, calculated in the pure molecular scenario for $X(3872)$ with allowance for the influence of the binding
of target nucleons and their Fermi motion on the direct processes (1), (2) and without it (cf. Eqs. (3) and (23))
is quite inessential at initial photon energies around energy of $E_{\gamma}=13$ GeV. This justifies the use of
approximate formulas (26) and (29) for calculation of the $X(3872)$ transparency ratios $S_A$ and $T_A$, respectively,
at least at this beam energy.

Figures 8 and 9 show the A--dependences of the transparency ratios $S_A$ and $T_A$
from the primary ${\gamma}N \to {X(3872)}N$ reaction channels in ${\gamma}A$
($A=$$^{12}$C, $^{27}$Al, $^{40}$Ca, $^{63}$Cu, $^{93}$Nb, $^{112}$Sn, $^{184}$W, $^{208}$Pb, and $^{238}$U)
collisions, calculated for the incident photon energy of around 13 GeV on the basis
of equations (26) and (29), correspondingly, and for the same
values of the genuine $X(3872)$--nucleon absorption cross section $\sigma_{X(3872)N}$
as those given above by Eq. (22).
One can see that the results  for the transparency ratio $S_A$ depend more strongly on this cross section
or on the adopted configuration for $X(3872)$ than those for the quantity $T_A$.
Thus, for the first observable, we observe the experimentally separated  differences $\sim$ 35, 25, 30, 10\%
between the results, obtained assuming for $X(3872)$ a pure charmonium and tetraquark scenarios,
a tetraquark and charmonium-molecule mixture with the charmonium and molecular probabilities of 50\% and 50\% pictures, a charmonium-molecule mixture with the charmonium and molecular probabilities of 50\% and 50\% and
a charmonium-molecule mixture with the charmonium and molecular probabilities of 15\% and 85\% configurations,
a charmonium-molecule mixture with the charmonium and molecular probabilities of 15\% and 85\% and pure molecular
pictures for comparatively "light" nuclei ($^{27}$Al, $^{40}$Ca).
For the medium-mass ($^{93}$Nb, $^{112}$Sn) and heavy ($^{184}$W, $^{238}$U) target nuclei these differences are
larger and are about 50, 35, 37, 11\% and 60, 41, 42, 12\%, respectively.
In the case of the quantity $T_A$ the analogous differences are somewhat smaller and they are about
10, 7, 6, 2\%, 25, 15, 12, 3\% and 40, 20, 17, 4\%, respectively, in the cases of "light", medium-mass
and heavy target nuclei indicated above.
We see yet that the highest sensitivity of both considered observables $S_A$ and $T_A$ to the $X(3872)$
structure is observed, as is expected, for heavy target nuclei.
Therefore, one can conclude that the observation, at least, of the A dependence of the transparency ratio $S_A$, especially for large mass numbers $A$, in the future CEBAF high-precision photoproduction experiments
\footnote{$^)$In which the data should be collected with experimental accuracy better than 10\%.}$^)$
offers the possibility to discriminate between all considered internal configurations of the $X(3872)$ exotic state.
On the other hand, the future precise $X(3872)$ photoproduction data on the A dependence of the transparency ratio $T_A$ in the range of large A, obtained in such experiments, could also additionally help to distinguish, at least, between a pure charmonium, tetraquark, charmonium-molecule mixture with the charmonium and molecular probabilities $\sim$ 50\% and 50\% and a pure molecular configurations of the $X(3872)$ resonance
\footnote{$^)$It should be noticed that such relative observable is favorable for the aim of getting the information on the $X(3872)$ internal structure from the experimental point of view, since it allows for a reduction of systematic errors due to the cancelation of the efficiency corrections.}$^)$
.

In addition, we have investigated the opportunity of discriminating between the proposed configurations for
the $X(3872)$ meson from the measurements of the incident photon energy dependence of the transparency ratio
$T_A$. Fig. 10 shows such dependence for the $^{184}$W/$^{12}$C combination.
It was calculated on the basis of Eq. (28) for five employed options for the $X(3872)$ meson inner structure and
for the off-shell intranuclear nucleons as well as in line with the simple formula (29) for its pure molecular
picture and for the free target nucleons being at rest. It can be seen from this figure that there is a definite
sensitivity of the transparency ratio $T_A$ to the considered variations in the $X(3872)$ meson--nucleon absorption
cross section or, which is equivalent, to its proposed configurations at all studied photon energies.
This sensitivity is similar to that available in Fig. 9 for heavy target nuclei. This means that this relative
observable can also be useful to help determine the $X(3872)$ inner structure. By looking at this figure, we see
yet that the simple expression (29) describes quite well the quantity $T_A$ at above threshold photon energies,
where it exhibits a practically flat behavior. But in the far subthreshold region ($E_{\gamma} \sim$ 8--10 GeV) this quantity increases as the photon-beam energy decreases. This can be explained by the fact that the averaged over
the Fermi motion and binding energy of the nucleons in the nucleus $X(3872)$ production cross sections calculated for the carbon nucleus decrease faster than the respective cross sections for the tungsten nucleus as the photon energy
becomes lower. This behavior of the transparency ratio $T_A$ can also be used for discriminating between possible
configurations of the $X(3872)$ state.

Finally, the absolute $X(3872)$ meson momentum distributions from the direct (1), (2) $X(3872)$ production processes
in $\gamma$$^{12}$C and $\gamma$$^{184}$W interactions, calculated on the basis of Eq. (32) for laboratory polar
angles of 0$^{\circ}$--10$^{\circ}$ and for incident photon energy of 13 GeV, are shown, respectively, in Figs. 11
and 12. These momentum distributions were obtained for five adopted values of the $X(3872)$--nucleon absorption
cross section (cf. Eq. (22)) in the considered scenarios for $X(3872)$. The absolute values of the differential cross
sections have a well measurable strength $\sim$ 10--10$^3$ nb/(GeV/c) in the high-momentum region of 11--12 GeV/c.
They also show a rather sizeable and experimentally distinguishable variations, especially for the heavy target nucleus
$^{184}$W
\footnote{$^)$Which are the same as those shown, respectively, in Figs. 4 and 5.}$^)$
, upon going over from the $X(3872)$--nucleon absorption cross section value of 3.5 to 42.0 mb. This behavior
of the differential cross sections can also be used to unveil the $X(3872)$ structure from comparison the present model
calculations with future experimental data.

   Taking into account the above considerations, we come to the conclusion that such observables as
the absolute total and differential cross sections for production of $X(3872)$ mesons from ${\gamma}A$ reactions
as well as their relative (transparency ratios) yields can be useful at near-threshold photon beam energies to help determine the genuine $X(3872)$ internal structure.

\section*{4. Epilogue}

\hspace{1.5cm} In this paper we have investigated the production of $X(3872)$ mesons in photon-induced nuclear
reactions near the threshold on the basis of an spectral function approach, which accounts for direct photon-nucleon $X(3872)$ production processes as well as five different scenarios for their internal structure.
We have calculated the absolute and relative excitation functions for the production of $X(3872)$ mesons off
$^{12}$C and $^{184}$W target nuclei at near-threshold incident photon energies of 8--16 GeV, the absolute differential cross sections for their production off these target nuclei at laboratory angles of 0$^{\circ}$--10$^{\circ}$ and for incident photon energy of 13 GeV as well as the A dependences of the
relative (transparency ratios) cross sections for $X(3872)$ production from ${\gamma}A$ collisions
at photon energies around 13 GeV within the adopted scenarios for the $X(3872)$ meson internal structure. We have shown that the absolute and relative observables considered reveal distinct sensitivity to these scenarios,
which implies that they may be an important tool to get valuable information on its inner configuration.
The measurements of these observables could be performed in the current and forthcoming experiments at the CEBAF facility [40--44], electron-ion colliders EIC [144] and EicC [145].
\\

\end{document}